\documentclass[12pt,a4paper]{article}
\usepackage[T1]{fontenc} 
\usepackage[dvipdfmx]{graphicx}
\usepackage{bm,caption,latexsym,amssymb,amsfonts,mathtools,here}
\usepackage{bm,latexsym,amsfonts,mathtools}
\usepackage{amssymb, amsmath, comment}
\usepackage{color}
\usepackage{hyperref}
\usepackage{booktabs}
\usepackage{mathrsfs}
\usepackage{wrapfig}
\usepackage{setspace}
\numberwithin{equation}{section}
\renewcommand{\baselinestretch}{1.5}

\usepackage[sorting=none,doi=false]{biblatex}

\bibliography{reference}

\begin{document}
\begin{titlepage}
\unitlength = 1mm
\begin{flushright}
KOBE-COSMO-22-08
\end{flushright}

\vskip 1cm
\begin{center}

{ \large \textbf{ Conversion of squeezed gravitons into photons during inflation}}

\vspace{1.8cm}
Sugumi Kanno$^*$, Jiro Soda$^{\flat}$, and Kazushige Ueda$^*$

\vspace{1cm}

\shortstack[l]
{\it $^*$ Department of Physics, Kyushu University, 744 Motooka, Nishi-ku, Fukuoka 819-0395, Japan \\ 
\it $^\flat$ Department of Physics, Kobe University, Kobe 657-8501, Japan
}

\vskip 4.0cm

{\large Abstract}\\
\end{center}

It is believed that relic gravitons are squeezed during inflation. If so,  quantum noise induced by them can be significantly enhanced in current interferometers. 
However, decoherence of the gravitons during cosmic history may change the degree of squeezing. As a first step for analyzing the decoherence of the gravitons, we assume the presence of a sizable magnetic field at the beginning of inflation and study conversion processes of the squeezed gravitons into photons during inflation in the case of minimal coupling between gravitons and photons.
We solve the dynamical evolution of a coupled system of gravitons and photons 
perturbatively and
obtain squeezing parameters for the system
 numerically and analytically.
It turns out that the gravitons are robust against
the decoherence caused by the cosmological magnetic fields.
We also find that the conversion rate of gravitons into photons is at a few percent at most. 
\vspace{1.0cm}
\end{titlepage}

\hrule height 0.075mm depth 0.075mm width 165mm
\tableofcontents
\vspace{1.0cm}
\hrule height 0.075mm depth 0.075mm width 165mm
\section{Introduction}
Inflation theory has claimed that the origin of the large-scale structure of the universe and temperature fluctuations in the cosmic microwave background radiations is quantum fluctuations. Remarkably, the inflation theory also predicts the existence of primordial gravitational waves stemming from the
quantum fluctuations of the spacetime (relic gravitons). After the discovery
of gravitational waves from a black hole binary system~\cite{LIGOScientific:2016aoc},
the detection of primordial gravitational waves has been
the most important research objective~\cite{Kawamura:2011zz,Amaro-Seoane:2012aqc}.

The notable nature  of primordial gravitational waves is their quantum origin. If the relic gravitons were found, it would strongly support the inflationary universe. The finding of the relic gravitons would also give a hint of quantum gravity. Hence, it is extremely important to explore
the quantum nature of the primordial gravitational waves.

It is well known that the generation of relic gravitons
can be interpreted as the squeezing process of a quantum state during inflation~\cite{Grishchuk:1989ss,Grishchuk:1990bj,Albrecht:1992kf,Polarski:1995jg}. Since the degree of squeezing is extremely high, the quantum state is highly entangled between two modes with opposite wave number vectors due to conservation of momentum. The squeezed state of the relic gravitons is a key for 
proving the nonclassicality of primordial gravitational waves. 
In fact, the squeezed gravitons can significantly enhance the quantum noise in interferometers~\cite{Parikh:2020nrd,Kanno:2020usf,Parikh:2020kfh,Parikh:2020fhy,Kanno:2021gpt}. 
Hence, we need to show the degree of the squeezing generated during inflation survives under the decoherence processes in the evolution of the universe. So far, the decoherence process due to short wavelength modes of a field has been investigated~\cite{Lombardo:2005iz,Martineau:2006ki,Burgess:2006jn,Nelson:2016kjm}. However, it is argued 
that the decoherence obtained by tracing out the short wavelength modes is false decoherence~\cite{2011arXiv1110.2199U}. 
Thus, it is worth studying different decoherence processes.

In this paper, as a source of the decoherence, we assume the presence of a sizable magnetic field at the beginning of inflation.  We then consider conversion process of the squeezed gravitons into photons during inflation in the case of minimal coupling between gravitons and photons~\cite{Gertsenshtein:1962,Raffelt:1987im,Chen:1994ch,Cillis:1996qy}. 
The squeezed state of gravitons may turn into the squeezed state of photons due to the graviton-photon conversion. 
Hence, it is important to clarify to what extent the squeezing of the relic gravitons survives at present. 
The purpose of this paper is to compute the 
degree of squeezing parameters of graviton and photon and cross squeezing parameter between gravitons and photons during inflation.

The paper is organized as follows: In section 2, we derive basic equations for analyzing the conversion process of gravitons into photons during inflation. 
In section 3, we explain the perturbative formalism for solving a coupled system between gravitons and photons in order to obtain
the time evolution of mode functions.
In section 4, we derive Bogoliubov transformations
due to the squeezing process in the presence of primordial magnetic fields. 
In section 5, we deduce formulae for the squeezing parameters and  
 reveal the time evolution of the squeezing parameters numerically and analytically. 
We also discuss the implications of our results.
The final section is devoted to the conclusion.

\section{Graviton-photon conversion during inflation}

We represent the graviton in a spatially flat expanding background by the tensor mode perturbation in the three-dimensional metric, 
\begin{eqnarray}
ds^2=a^2(\eta)\left[-d\eta^2+\left(\delta_{ij}+h_{ij}\right)dx^idx^j\right]\,,
\end{eqnarray}
where $\eta$ is the conformal time and the metric perturbation $h_{ij}$ satisfies the transverse traceless conditions $h_{ij}{}^{,j}=h^i{}_i=0$. The spatial indices $i,j,k,\cdots$ are raised and lowered by $\delta^{ij}$ and $\delta_{k\ell}$.

The Einstein-Hilbert action and the action for the electromagnetic field is given by
\begin{eqnarray}
S=S_g+S_A=\frac{M_{\rm pl}^2}{2}\int d^4x \sqrt{-g}
\,
R-\frac{1}{4}\int d^4x \sqrt{-g}\,
F^{\mu\nu}
F_{\mu\nu}
\label{original action}\,,
\end{eqnarray}
where $M_{\rm pl}=1/\sqrt{8\pi G}$ is the Planck mass. The gauge field $A_\mu$ represents the photon and the field strength is defined by $F_{\mu\nu}=\partial_\mu A_{\nu}-\partial_\nu A_{\mu}$.
Expanding the Einstein-Hilbert action up to the second order in perturbations $h_{ij}$, we find
\begin{eqnarray}
\delta S_g=\frac{M_{\rm pl}^2}{8}\int d^4x\,a^2\left[
h^{ij\prime}\,h_{ij}^\prime-h^{ij,k}h_{ij,k}
\right]\,.
\label{action:g}
\end{eqnarray}
Here, a prime denotes the derivative with respect to the conformal time. The action for the photon up to second order in perturbations $A_i$ reads
\begin{eqnarray}
\delta S_A=\frac{1}{2}\int d^4x\left[A_i^{\prime\, 2}-A_{k,i}^2\right]\,,
\label{action:A}
\end{eqnarray}
where the photon field satisfies the Coulomb gauge $A_0=0$ and $A^i{}_{,i}=0$.
The action for the interaction between the graviton and the photon  up to second order in perturbations $h_{ij}, A^i$ is found to be
\begin{eqnarray}
\delta S_{\rm I}=\int d^4x \left[
\varepsilon_{i\ell m}B_m h^{ij}\left(\partial_j A_\ell
-\partial_\ell A_j\right)
\right]\,.
\label{action:I}
\end{eqnarray}
Note that $B_m=\varepsilon_{mj\ell}\,\partial_j A_\ell$ is a constant background magnetic field that we assumed the presence at the beginning of inflation.

At quadratic order, it is convenient to expand $h_{ij}(\eta,x^i)$ and $A_i(\eta,x^i)$ in the Fourier modes,
\begin{align}
&h_{ij}(\eta,x^i)=\frac{2}{M_{\rm pl}} \sum_{P}\frac{1}{(2\pi)^{3/2}} \int d^3 k\,h^{P}_{\bm k}(\eta)\, e_{ij}^{P}(\bm{k})\,e^{i\bm{k}\cdot\bm{x}}\,,\\
&A_i(\eta,x^i)=\sum_{P} \frac{\pm i}{(2\pi)^{3/2}}
\int d^3 k\,A^{P}_{\bm k}(\eta)\,e_i^{P}(\bm{k})\, e^{i\bm{k}\cdot\bm{x}},
\end{align}
where three-vectors are denoted by bold math type and  $e_{ij}^{P}(\bm{k})$ and $e_i^{P}(\bm{k})$ are the polarization tensors and vectors for the ${\bm k}$ mode respectively normalized as $e^{ijP}(\bm{k})e_{ij}^{Q}(\bm{k})=\delta^{PQ}$ and $e^{iP}(\bm{k}) e_i^{Q}(\bm{k})=\delta^{PQ}$ with $P,Q=+,\times$.
Using the canonical variable $y^P_{\bm k}(\eta)=a(\eta)h_{\bm k}^{P}(\eta)$,
we can rewrite the quadratic actions~(\ref{action:g}), (\ref{action:A}) and (\ref{action:I})  as
\begin{eqnarray}
\delta S_g&=&\frac{1}{2}\sum_P\int d^3 k\,d\eta\left[\,
|y_{\bm k}^{P\,\prime}|^2
-k^2|y_{\bm k}^P|^2
-\frac{a^\prime}{a}y_{\bm k}^{P}y_{-\bm k}^{P\,\prime}
-\frac{a^\prime}{a}y_{-\bm k}^{P}y_{\bm k}^{P\,\prime}
+\left(\frac{a^\prime}{a}\right)^2|y_{\bm k}^P|^2
\right]\,,
\label{action:g2}\\\
\delta S_A&=&\frac{1}{2}\sum_P\int d^3 k\,d\eta\left[\,
|A_{\bm k}^{P\,\prime}|^2-k^2|A_{\bm k}^P|^2
\,\right]\,,
\label{action:A2}\\
\delta S_I&=&\frac{2}{M_{\rm pl}}\sum_{P,Q}\int d^3 k\,d\eta\,\frac{1}{a}\left[
\varepsilon_{i\ell m}\,B_m\,y_{\bm k}^PA_{-\bm k}^Q
\,e_{ij}^P(\bm k)\Bigl\{ik_\ell\,e_{j}^Q(-\bm k)-ik_j\,e_{\ell }^Q(-\bm k)\Bigr\}\right]
\label{action:I2}\,,
\end{eqnarray}
where $k=|\bm k|$. Polarization vectors $e^{i+}, e^{i\times}$ and a vector $k^i/k$ constitute an orthnormal basis.
Without loss of generality, we assume the constant background magnetic field is in the ($k^i, e^{i \times}$)-plane.
\begin{figure}[H]
\centering
 \includegraphics[keepaspectratio, scale=0.6]{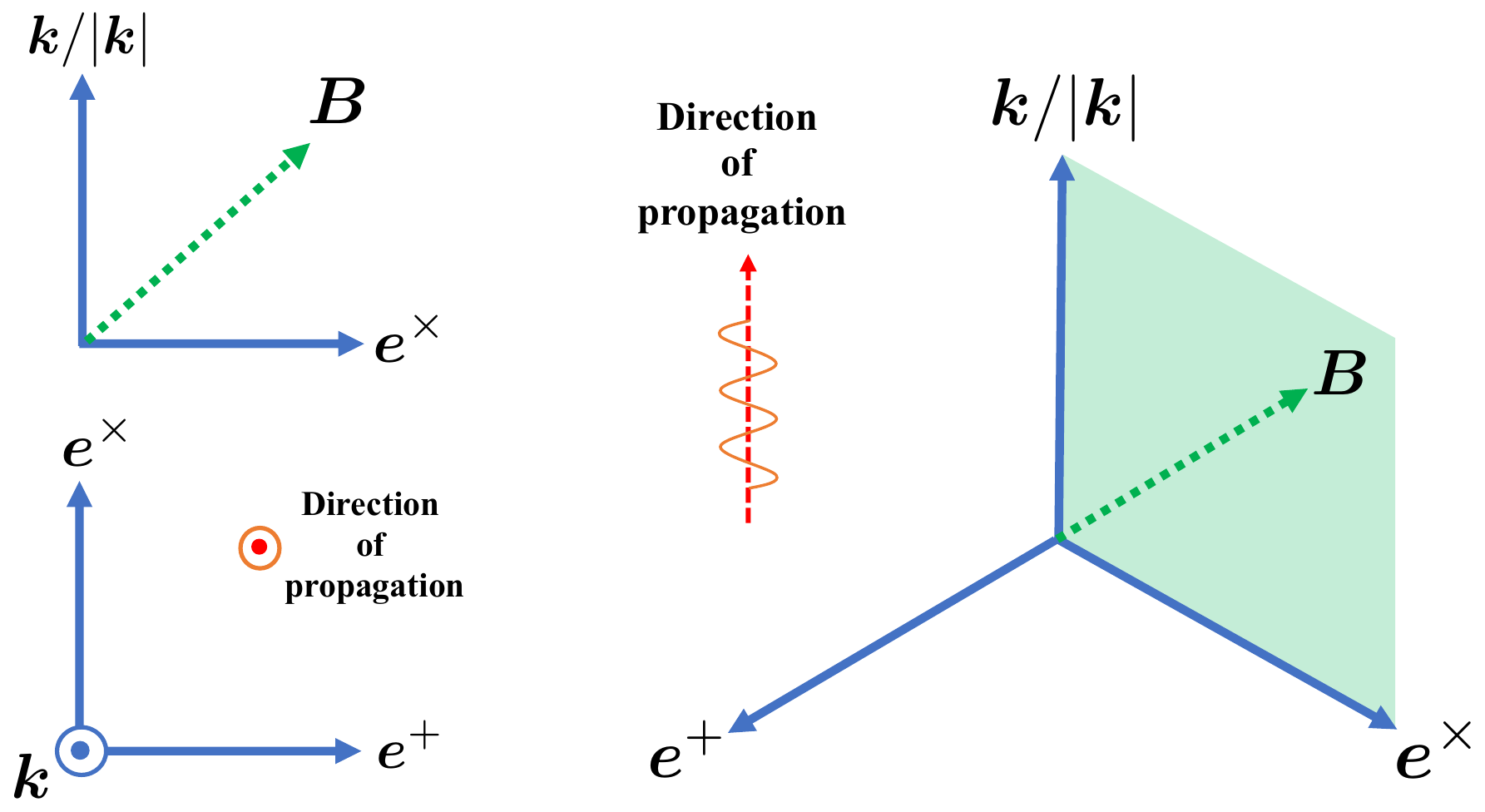}
 \renewcommand{\baselinestretch}{3}
 \caption{Configuration of the polarization vector ${\bm e}^P(\bm k)$, wave number ${\bm k}$, and background magnetic field ${\bm B}$.}
 \label{Configuration}
 \end{figure}
\noindent
The polarization tensors  can be written in terms of polarization vectors  
$e^{i+}$ and $e^{i\times}$
 as
\begin{align}
    &e_{ij}^+(\bm{k})
    =\frac{1}{\sqrt{2}} \Bigl\{
    e^+_i(\bm{k}) e^+_j(\bm{k})-e^\times_i(\bm{k}) e^\times_j(\bm{k})
    \Bigr\}\,,\\
    &e_{ij}^\times(\bm{k})
    =\frac{1}{\sqrt{2}} 
    \Bigl\{
    e^+_i(\bm{k}) e^\times_j(\bm{k})+e^\times_i(\bm{k}) e^+_j(\bm{k})
    \Bigr\}\, .
\end{align}
In the following, we assume
\begin{eqnarray}
e_i^\times(-\bm{k})=-e_i^\times(\bm{k})\,.
\end{eqnarray}
The action (\ref{action:I2}) is then reduced into
\begin{eqnarray}
\delta S_I&=&\frac{\sqrt{2}}{M_{\rm pl}}\int d^3k\,d\eta\,\frac{1}{a}
\left[\,\lambda(\bm k)\,y_{\bm k}^+(\eta)\,A_{-\bm k}^+(\eta)+\lambda(\bm k)\,y_{\bm k}^\times(\eta)\,A_{-\bm k}^\times(\eta)\,\right]\,,
\label{action:I3}
\end{eqnarray}
where we defined the coupling between graviton and photon as 
\begin{align}
\lambda(\bm{k})\equiv\frac{\sqrt{2}}{M_{\rm pl}}
\varepsilon^{i\ell m}\,e_i^+\,k_\ell\,B_m\,.
\label{coupling}
\end{align}
Here, the conditions for the graviton and photon to be real read,
$h_{-\bm k}^{+,\times}(\eta)=h_{\bm k}^{*\,+,\times}(\eta)$ and $A_{-\bm k}^{+,\times}(\eta)=-A_{\bm k}^{*\,+,\times}(\eta)$\,.
Below, we focus on the plus polarization and omit the index $P$ unless there may be any confusion. 

In the case of de Sitter space, the scale factor is given by $a(\eta)=-1/(H\eta)$ where $-\infty<\eta<0$.
The variation of the actions (\ref{action:g2}), (\ref{action:A2}) and (\ref{action:I3}) with respect to the graviton and the photon fields gives
\begin{align}
    &y_{\bm{k}}''+\left(k^2-\frac{2}{\eta^2}\right)y_{\bm{k}}=\lambda H \eta A_{\bm{k}}\,,
    \label{eom:graviton}
    \\
    &A_{\bm{k}}''+k^2A_{\bm{k}}=\lambda H \eta\,y_{\bm{k}}\,.
    \label{eom:photon}
\end{align}
If we define the Lagrangian in the actions (\ref{action:g2}) and (\ref{action:A2}) by $\delta S_g=\int d\eta\,L_g$ and $\delta S_A=\int d\eta\,L_A$, the conjugate momenta of graviton $p_{\bm k}$ and photon $\pi_{\bm k}$ are respectively given by
\begin{align}
    &p_{\bm{k}}(\eta)=\frac{\partial L_g}{\partial y^\prime_{-\bm k}}=y_{\bm{k}}'(\eta)+\frac{1}{\eta}y_{\bm{k}}(\eta) \ , 
    \label{p}\\
    &\pi_{\bm{k}}(\eta)=\frac{\partial L_A}{\partial A^\prime_{-\bm k}}=A_{\bm{k}}'(\eta) \ .
    \label{pi}
\end{align}
Now we promote variables $y_{\bm k}(\eta), A_{\bm k}(\eta)$ and their momenta $p_{\bm k}(\eta), \pi_{\bm k}(\eta)$ into operators. The annihilation operator for the graviton is expressed by canonical variables as
\begin{eqnarray}
\hat{a}_y(\eta,{\bm k})=\sqrt{\frac{k}{2}}\hat{y}_{\bm k}(\eta)+\frac{i}{\sqrt{2k}}\hat{p}_{\bm k}(\eta)\,.
\label{y:annihi}
\end{eqnarray}
In the same way, the annihilation operator for photon is given by
\begin{eqnarray}
\hat{a}_A(\eta,{\bm k})=\sqrt{\frac{k}{2}}\hat{A}_{\bm k}(\eta)+\frac{i}{\sqrt{2k}}\hat{\pi}_{\bm k}(\eta)\,.
\label{A:annihi}
\end{eqnarray}
The commutation relations $[\hat{a}_y(\eta,{\bm k}),\hat{a}^\dag_y(\eta,-{\bm k}^\prime)]=\delta({\bm k}+{\bm k}^\prime)$ and $[\hat{a}_A(\eta,{\bm k}),\hat{a}^\dag_A(\eta,-{\bm k}^\prime)]=\delta({\bm k}+{\bm k}^\prime)$ guarantee the canonical commutation relations $[y_{\bm k}(\eta),p_{{\bm k}^\prime}(\eta)]=i\delta({\bm k}-{\bm k}^\prime)$ and $[A_{\bm k}(\eta),\pi_{{\bm k}^\prime}(\eta)]=i\delta({\bm k}-{\bm k}^\prime)$.
Notice that the annihilation operator becomes time dependent through the time dependence of canonical variables. Thus, the vacuum defined by $\hat{a}(\eta,{\bm k})|0\rangle =0$ is time dependent as well and the vacuum in this formalism turns out to be defined at every moment.


In this paper, we suppose $B_m/M_{\rm pl}\ll 1$ so that the coupling between graviton and photon~(\ref{coupling}) is weak . Then we solve the Eqs.~(\ref{eom:graviton}) and (\ref{eom:photon}) iteratively up to the second order in $y_{\bm{k}}$ and $A_{\bm{k}}$ in the next section.

\section{Time evolution of mode functions}

Using the basic equations presented in the previous section, we perturbatively derive mode functions in this section.

\subsection{Zeroth order}

By letting $\lambda=0$ in Eqs.~(\ref{eom:graviton}) and (\ref{eom:photon}), the equations of the zeroth order approximation become
\begin{align}
    &\hat{y}_{\bm{k}}^{(0)\prime\prime}+\left(k^2-\frac{2}{\eta^2}\right)\hat{y}_{\bm{k}}^{(0)}=0\,,
    \label{GWeqs}
    \\
    &\hat{A}_{\bm{k}}^{(0)\prime\prime}+k^2\hat{A}_{\bm{k}}^{(0)}=0 \, ,
    \label{EMeqs}
\end{align}
where the superscript $(0)$ denotes the zeroth order. The solutions of the above equations are
\begin{align}
    &\hat{y}_{\bm k}^{(0)}(\eta)=u_{\bm k}^{(0)}(\eta) ~\hat{c}
    +u_{\bm k}^{(0)*}(\eta)~\hat{c}^\dagger\,, 
    \label{0th:graviton}\\
    &\hat{A}_{\bm k}^{(0)}(\eta)=v_{\bm k}^{(0)}(\eta) ~\hat{d}
    +v_{\bm k}^{(0)*}(\eta)~\hat{d}^\dagger,
    \label{0th:photon}
\end{align}
where $\hat{c}$\,($\hat{d}$) and its conjugate 
$\hat{c}^\dag$($\hat{d}^\dag$) are constant operators of integration. We choose the properly normalized positive frequency mode in the remote past as a basis, which is expressed as
\begin{align}
    &u_{\bm k}^{(0)}(\eta)=\frac{1}{\sqrt{2k}} \biggl(1-\frac{i}{k\eta}\biggr) e^{-ik\eta}
    \,,\qquad
    v_{\bm k}^{(0)}(\eta)=\frac{1}{\sqrt{2k}} e^{-ik\eta}.
\end{align}

\subsection{First order}

Inserting the solutions of zeroth order approximation (\ref{0th:graviton}) and (\ref{0th:photon}) into the r.h.s of Eqs.~(\ref{eom:graviton}) and (\ref{eom:photon}) as the source terms, the equations of the first order approximation  are written as
\begin{align}
    &\hat{y}_{\bm{k}}^{(1)\prime\prime}+\left(k^2-\frac{2}{\eta^2}\right)\hat{y}_{\bm{k}}^{(1)}
    =\lambda H \eta \hat{A}_{\bm{k}}^{(0)}\,,
    \label{GWeq1}
    \\
    &\hat{A}_{\bm{k}}^{(1)\prime\prime}+k^2\hat{A}_{\bm{k}}^{(1)}=\lambda H \eta \hat{y}_{\bm{k}}^{(0)} \,.
    \label{EMeq1}
\end{align}
The effect of photon comes in Eq.~(\ref{GWeq1}). Using the Green function
\begin{eqnarray}
G_{\rm dS}(\eta,\eta')=
\frac{1}{2ik} \biggl(1+\frac{i}{k\eta'}\biggr)
\biggl(1-\frac{i}{k\eta}\biggr)
e^{-ik(\eta-\eta')}
-\frac{1}{2ik} 
\biggl(1-\frac{i}{k\eta'}\biggr)
\biggl(1+\frac{i}{k\eta}\biggr)
e^{ik(\eta-\eta')}\,,
\end{eqnarray}
we obtain the solution  as
\begin{align}
    \hat{y}^{(1)}_{\bm k}(\eta)
    &=-\int_{\eta_i}^\eta d\eta'
    G_{\rm dS} (\eta,\eta')
    \lambda H \eta'\hat{A}^{(0)}_{\bm k}(\eta')\nonumber\\
    &=-\int_{\eta_i}^\eta d\eta'
    G_{\rm dS} (\eta,\eta')
    \lambda H \eta' 
    v^{(0)}_{\bm k}(\eta')~\hat{d}
    -\int_{\eta_i}^\eta d\eta'
    G_{\rm dS} (\eta,\eta')
    \lambda H\eta'
    v^{(0)*}_{\bm k}(\eta') 
    ~\hat{d}^\dagger \nonumber\\
    &\equiv u^{(1)}_{\bm k}(\eta)~\hat{d}+u^{(1)*}_{\bm k}(\eta)~\hat{d}^\dagger\,,
    \label{1st:graviton}
\end{align}
where $\eta_i$ is an initial time. From the first line to the second line we used Eq.~(\ref{0th:photon}). In the last line, we defined the first order correction due to the source of photon to the positive frequency mode of graviton by
\begin{align}
    u^{(1)}_{\bm k}(\eta)&\equiv-\int_{\eta_i}^\eta d\eta'
    G_{\rm dS} (\eta,\eta')
    \lambda H \eta' 
    v^{(0)}_{\bm k}(\eta') \ .
    \label{u1Green}
\end{align}
After integration, we have
\begin{eqnarray}
u^{(1)}_{\bm k}(\eta)&=&
\frac{ \lambda  H }{8 \sqrt{2} \eta  k^{9/2}} 
\left[e^{- i k \eta } 
\Bigl\{2 i \eta ^3 k^3+\eta  k 
\Bigl(2 \eta_i k (2-i \eta_i k)+3 i\Bigr)
-2 \eta_i k 
(\eta_i k+2 i)
+3\Bigr\}
\right.\nonumber\\
&&\left.
-e^{ i k (\eta -2\eta_i )  } 
(\eta  k+i) (2 \eta_i k-3 i)
\right]  \ .
\label{u1}
\end{eqnarray}

Similarly, the effect of graviton comes in Eq.~(\ref{EMeq1}).
By using the Green function 
\begin{eqnarray}
G_{\rm M}(\eta,\eta')=-\frac{1}{k} \sin{k(\eta-\eta')}\,,
\end{eqnarray}
we have
\begin{align}
    \hat{A}^{(1)}_{\bm k}(\eta)
    &=-\int_{\eta_i}^\eta d\eta'
    G_{\rm M} (\eta,\eta')
    \lambda H \eta' \hat{y}^{(0)}_{\bm k}(\eta')\nonumber\\
    &=-\int_{\eta_i}^\eta d\eta' G_{\rm M}(\eta,\eta')
    \lambda H \eta' u_{\bm k}^{(0)}(\eta)~\hat{c}
    -\int_{\eta_i}^\eta d\eta' G_{\rm M}(\eta,\eta')
    \lambda H \eta' u_{\bm k}^{(0)*}(\eta)~\hat{c^\dagger}\nonumber\\
    &\equiv v^{(1)}_{\bm k}(\eta)~\hat{c}+v^{(1)*}_{\bm k}(\eta)~\hat{c}^\dagger \ ,
    \label{1st:photon}
\end{align}
where we used Eq.~(\ref{0th:graviton}) from the first line to the second line. We also defined the first order correction due to the source of graviton to the positive frequency mode of photon in the third line by
\begin{align}
    v^{(1)}_{\bm k}(\eta)&\equiv-\int_{\eta_i}^\eta d\eta'
    G_{\rm M} (\eta,\eta')
    \lambda H \eta' 
    u^{(0)}_{\bm k}(\eta') \ .
    \label{v1Green}
\end{align}
More explicitly, the above is written as
\begin{eqnarray}
v^{(1)}_{\bm k}(\eta)
&=&
\frac{ \lambda H }{8 \sqrt{2} k^{7/2}}
\left[e^{- i k \eta} \Bigl\{2 i k^2 (\eta^2 -\eta_i^2) +k (6 \eta -4 \eta_i)-3 i\Bigr\}
+e^{ i k(\eta -2\eta_i )  } (-2 \eta_i k+3 i)\right] \ .\nonumber\\
\label{v1}   
\end{eqnarray}

\subsection{Second order}

By plugging the solution of the first order approximation (\ref{1st:graviton}) and (\ref{1st:photon}) into the r.h.s of Eqs.~(\ref{eom:graviton}) and (\ref{eom:photon}) as the source terms, the equations of the second order approximation are
\begin{align}
    &y_{\bm{k}}^{(2)\prime\prime}+\left(k^2-\frac{2}{\eta^2}\right)y_{\bm{k}}^{(2)}
    =\lambda H \eta A_{\bm{k}}^{(1)}\,,
    \label{GWeq2}
    \\
    &A_{\bm{k}}^{(2)\prime\prime}+k^2A_{\bm{k}}^{(2)}=\lambda H \eta\,y_{\bm{k}}^{(1)} \,.
    \label{EMeq2}
\end{align}
At this order, the effect of graviton itself comes in Eq.~(\ref{GWeq2}). The solution is written by the Green function $G_{\rm dS}$ such as
\begin{align}
    \hat{y}^{(2)}_{\bm k}(\eta)
    &=-\int_{\eta_i}^\eta d\eta'
    G_{\rm dS} (\eta,\eta')
    \lambda H \eta' \hat{A}^{(1)}_{\bm k} (\eta') \nonumber\\
    &=-\int_{\eta_i}^\eta d\eta'
    G_{\rm dS} (\eta,\eta')
    \lambda H \eta' v^{(1)}_{\bm k}(\eta')~\hat{c}
    -\int_{\eta_i}^\eta d\eta'
    G_{\rm dS} (\eta,\eta')
    \lambda H \eta' v^{(1)*}_{\bm k}(\eta')~\hat{c}^\dagger
    \nonumber\\
    &=-\int_{\eta_i}^{\eta} d\eta' G_{\rm dS} (\eta,\eta')
    \lambda H \eta' \Biggl(-\int_{\eta_i}^{\eta'}d\eta'' G_{\rm M}(\eta',\eta'')\lambda H \eta''u^{(0)}_{\bm k}(\eta'') \Biggr)~ \hat{c} \nonumber\\
    &\qquad -\int_{\eta_i}^{\eta} d\eta' G_{\rm dS} (\eta,\eta')
    \lambda H \eta' \Biggl(-\int_{\eta_i}^{\eta'}d\eta''G_{\rm M}(\eta',\eta'') \lambda H \eta''u^{(0)*}_{\bm k}(\eta'') \Biggr)~ \hat{c}^\dagger \nonumber\\
    &\equiv u^{(2)}_{\bm k}(\eta)~\hat{c}+u^{(2)*}_{\bm k}(\eta)~\hat{c}^\dagger, 
\end{align}
where we used Eqs.~(\ref{1st:photon}) and (\ref{v1Green}) in the second and the third lines respectively. In the last line, we defined
\begin{align}
    u^{(2)}_{\bm k}(\eta)
    &\equiv
    -\int_{\eta_i}^\eta d\eta' G_{\rm dS} (\eta,\eta')
    \lambda H \eta'
    v^{(1)}_{\bm k}(\eta')
    \nonumber\\
    &=\int_{\eta_i}^\eta d\eta' G_{\rm dS} (\eta,\eta')
    \lambda H \eta'
    \int_{\eta_i}^{\eta'} d\eta'' G_{\rm M}(\eta',\eta'') 
    \lambda H \eta''  u^{(0)}_{\bm k} (\eta'').
    \label{u2Green}
\end{align}
By performing the integration, the explicit form of the $u^{(2)}_{\bm k}(\eta)$ is found to be
\begin{eqnarray}
u^{(2)}_{\bm k}(\eta)
&=&
-\frac{ \lambda^2 H^2}{192 \sqrt{2} \eta  k^{15/2}}\nonumber\\
&&\times
\Biggl[
3 e^{ i k ( \eta -2\eta_i)} 
\biggl\{2 \eta^3 k^3 (-3-2 i \eta_i k)+\eta  k \biggl(-35+2 \eta_i k (\eta_i k (11+2 i \eta_i k)-23 i)\bigg)\nonumber\\
&&  \qquad  +2 \eta_i k (23+\eta_i k (-2 \eta_i k+11 i))-35 i
\biggr\}\nonumber\\
&&
+e^{- i k \eta }
\Biggl\{k 
\Bigl(-105 \eta +72 \eta_i+6 \eta  k^4 \left(\eta^2-\eta_i^2\right)^2-2 i k^3 \left(17 \eta^4-12 \eta^3 \eta_i-8 \eta\eta_i^3+3 \eta_i^4\right)\nonumber\\
&& \qquad +k^2 \left(16 \eta_i^3-52 \eta^3\right)+72 i \eta\eta_i k
\Bigr)
+105 i
\Biggr\}
\Biggr] \ .
\label{u2}
\end{eqnarray}
Similarly, the effect of photon itself comes in Eq.~(\ref{EMeq2}) and the solution is given by
\begin{align}
    \hat{A}^{(2)}_{\bm k}(\eta)
    &=-\int_{\eta_i}^\eta d\eta' G_{\rm M} (\eta,\eta')
    \lambda H \eta'~\hat{y}^{(1)}_{\bm k} (\eta') \nonumber\\
    &=-\int_{\eta_i}^\eta d\eta' G_{\rm M} (\eta,\eta')
    \lambda H \eta'u^{(1)}_{\bm k}(\eta')~\hat{d} 
    -\int_{\eta_i}^\eta d\eta' G_{\rm M} (\eta,\eta')
    \lambda H \eta'~u^{(1)*}_{\bm k}(\eta')~\hat{d}^\dagger \nonumber\\
    &=-\int_{\eta_i}^\eta d\eta' G_{\rm M}(\eta,\eta')
    \lambda H \eta'~
    \Biggl( 
    -\int_{\eta_i}^{\eta'} d\eta'' G_{\rm dS}(\eta',\eta'') 
    \lambda H \eta''  v^{(0)}_{\bm k} (\eta'')
    \Biggr) \hat{d}
    \nonumber\\
    &\qquad-\int_{\eta_i}^\eta d\eta' G_{\rm M} (\eta,\eta')
    \lambda H \eta'~
    \Biggl( 
    -\int_{\eta_i}^{\eta'} d\eta'' G_{\rm dS}(\eta',\eta'') 
    \lambda H \eta''  v^{(0)*}_{\bm k} (\eta'')
    \Biggr) \hat{d}^\dagger
    \nonumber\\
    &=v^{(2)}_{\bm k}(\eta)~\hat{d}+v^{(2)*}_{\bm k}(\eta)~\hat{d}^\dagger,
\end{align}
where we used Eqs.~(\ref{1st:graviton}) in the second line and (\ref{u1Green}) in the third line and in the last line. We defined
\begin{align}
    v^{(2)}_{\bm k}(\eta)
    &\equiv
    -\int_{\eta_i}^\eta d\eta' G_{\rm M} (\eta,\eta')
    \lambda H \eta'
    u^{(1)}_{\bm k}(\eta')
    \nonumber\\&=\int_{\eta_i}^\eta d\eta' G_{\rm M}(\eta,\eta')
    \lambda H \eta'
    \int_{\eta_i}^{\eta'} d\eta'' G_{\rm dS}(\eta',\eta'') 
    \lambda H \eta''  v^{(0)}_{\bm k} (\eta'')\,.
    \label{v2Green}
\end{align}
The integral of the above reduces to
\begin{eqnarray}
v^{(2)}_{\bm k}(\eta)
&=&
-\frac{ \lambda^2 H^2 }{64 \sqrt{2} k^{13/2}}\nonumber\\
&&\times
\Biggl[
e^{- i k \eta } 
\Biggl(2 k^4 \left(\eta^2-\eta_i^2\right)^2-4 i \eta  k^3 (\eta -\eta_i) (\eta +3 \eta_i)\nonumber\\
&& \qquad  +12 \eta_i k^2 (\eta_i-2 \eta )-12 i k (\eta -2 \eta_i)-15
\Biggr)\nonumber\\
&&\qquad +e^{ i k ( \eta -2 \eta_i)} 
\left(2 k (3+2 i \eta_i k) \left(\eta_i^2 k-\eta  (\eta  k+3 i)\right)+6 i \eta_i k+15
\right)
\Biggr] \ .
\label{v2}
\end{eqnarray}

\section{Bogoliubov transformations}

By solving Eqs.(\ref{eom:graviton}) and (\ref{eom:photon}) iteratively up to the second order, we can take into account the backreaction of graviton and photon respectively.  For the graviton,
the field and its conjugate momentum are now given by
\begin{eqnarray}
\hspace{-1cm}
\hat{y}_{\bm k}(\eta)&=&\Bigl(u^{(0)}_{\bm k}+u^{(2)}_{\bm k}\Bigr)\,\hat{c}
+u^{(1)}_{\bm k}\,\hat{d}
+{\rm h.c.} \,,
\\
\label{pGWsol}
\hat{p}_{\bm k}(\eta)&=&\Bigl(u^{(0)\,\prime}_{\bm k}+u^{(2)\,\prime}_{\bm k}\Bigr)\,\hat{c}
+u^{(1)\,\prime}_{\bm k}\,\hat{d}
+\frac{1}{\eta}\Bigl\{\left(
u^{(0)}_{\bm k}+u^{(2)}_{\bm k}
\right)\hat{c}+u^{(1)}_{\bm k}\,\hat{d}\Bigr\}+{\rm h.c.}\,,
\end{eqnarray}
where we used Eq.~(\ref{p}) and h.c. represents Hermitian conjugate. For the photon,
the field and its conjugate momentum become
\begin{eqnarray}
\hat{A}_{\bm k}(\eta)&=&\Bigl(v^{(0)}_{\bm k}+v^{(2)}_{\bm k}\Bigr)\,\hat{d}
+v^{(1)}_{\bm k}\,\hat{c}
+{\rm h.c.}\,,\\
\hat{\pi}_{\bm k}(\eta)&=&
\Bigl(v^{(0)\,\prime}_{\bm k}+v^{(2)\,\prime}_{\bm k}\Bigr)\,\hat{d}
+v^{(1)\,\prime}_{\bm k}\,\hat{c}
+{\rm h.c.}\,,
\label{pEMsol}
\end{eqnarray}
where we used Eq.~(\ref{pi}).
Then the annihilation operators for the graviton and photon are obtained by using Eqs.~(\ref{y:annihi}) and (\ref{A:annihi}) such as
\begin{align}
    \hat{a}_{y}(\eta,\bm{k})
    &= \Bigl(
    \psi_{p}^{(0)}
    +\psi_{p}^{(2)}
    \Bigr)\hat{c}
    +\Bigl(
    \psi_{m}^{(0)*}
    +\psi_{m}^{(2)*}
    \Bigr)\hat{c}^\dagger
    +\psi_{ p}^{(1)} \hat{d}
    +\psi_{m}^{(1)*} \hat{d}^\dagger\,,
    \label{y:annihifull}
    \\
    \hat{a}_{A}(\eta,\bm{k})
    &= \Bigl(
    \phi_{p}^{(0)}
    +\phi_{p}^{(2)}
    \Bigr)\hat{d}
    +\Bigl(
    \phi_{ m}^{(0)*}
    +\phi_{ m}^{(2)*}
    \Bigr)\hat{d}^\dagger
    +\phi_{ p}^{(1)} \hat{c}
    +\phi_{ m}^{(1)*} \hat{c}^\dagger\,.
    \label{A:annihifull}
\end{align}
Here, we defined new variables
\begin{align}
&\psi_{p}^{(j)} 
    =\sqrt{\frac{k}{2}} u^{(j)}_{\bm k}(\eta)
    +\frac{i}{\sqrt{2k}}
    \Bigl(u^{(j)\prime}_{\bm k}(\eta)+\frac{1}{\eta}u^{(j)}_{\bm k} (\eta)
    \Bigr) ,\label{psip}\\
&\psi_{ m}^{(j)}
    =\sqrt{\frac{k}{2}} u^{(j)}_{\bm k}(\eta)
    -\frac{i}{\sqrt{2k}}
    \Bigl(u^{(j)\prime}_{\bm k}(\eta)+\frac{1}{\eta}u^{(j)}_{\bm k}(\eta)
    \Bigr),  \label{psim}\\
&\phi_{ p}^{(j)}
    =\sqrt{\frac{k}{2}} v^{(j)}_{\bm k}(\eta)
    +\frac{i}{\sqrt{2k}} v_{\bm k}^{(j)\prime}(\eta),
    \label{phip}\\
&\phi_{ m}^{(j)} 
    =\sqrt{\frac{k}{2}} v^{(j)}_{\bm k}(\eta)
    -\frac{i}{\sqrt{2k}}  v_{\bm k}^{(j)\prime}(\eta),
    \label{phim}
\end{align}
where $j=0,1,2$ denotes the order of perturbations.

We see that all mode functions other than the zeroth order given in Eqs.~(\ref{u1Green}), (\ref{v1Green}) (\ref{u2Green}) and (\ref{v2Green}) vanish at the initial time $\eta_i$. Thus only the zeroth order of the above Eqs.~(\ref{psip}) $\sim$ (\ref{phim}) remains at the initial time. This means that annihilation operators in Eqs.(\ref{y:annihifull}) and (\ref{A:annihifull}) at the initial time are expressed by the zeroth order variables
\begin{align}
    \hat{a}_{y}(\eta_i,\bm{k})
    &=\left(1-\frac{i}{2k\eta_i}\right)e^{-ik\eta_i}\,\hat{c}
    +\frac{i}{2k\eta_i} e^{ik\eta_i}\,\hat{c}^\dagger,
    \label{ycRel}
    \\
    \hat{a}_A(\eta_i,\bm{k})
    &=e^{-ik \eta_i} \hat{d}.
\label{AdRel}
\end{align}
Combining Eqs. (\ref{ycRel}) and (\ref{AdRel}) with their complex conjugate, we can express the  $\hat{c}$ and $\hat{d}$  by the initial creation and annihilation 
operators as
\begin{eqnarray}
    \hat{c}  &=& \left( 1+\frac{i}{2k\eta_i}\right)e^{ik\eta_i}\,\hat{a}_y(\eta_i,\bm{k} )
    -\frac{i}{2k\eta_i} e^{ik\eta_i}\,\hat{a}_y^\dagger (\eta_i,-\bm{k}) \ , \\
  \hat{d}  &=&   e^{ik\eta_i}\,\hat{a}_A(\eta_i,\bm{k})  \ .
\end{eqnarray}
Plugging the above back into Eqs.(\ref{y:annihifull}) and (\ref{A:annihifull}), the time evolution of annihilation operator of graviton is described by the Bogoliubov transformation in the form
\begin{align}
    &\hat{a}_y(\eta,\bm{k})=
    \Biggl[
    \biggl(
    \psi_{p}^{(0)}
    +\psi_{p}^{(2)}
    \biggr)
    \Bigl( 1+\frac{i}{2k\eta_i} \Bigr)
    e^{ik\eta_i} 
    +\biggl(
    \psi_{m}^{(0)*}
    +\psi_{ m}^{(2)*}
    \biggr) \frac{i}{2k\eta_i} e^{-ik\eta_i} 
    \Biggr]
    \hat{a}_y(\eta_i,\bm{k})\nonumber\\
    &\hspace{1.5cm}
    +\Biggl[
    \biggl(
    \psi_{p}^{(0)}
    +\psi_{p}^{(2)}
    \biggr)\Bigl(-\frac{i}{2k\eta_i} \Bigr) e^{ik\eta_i} 
    +\biggl(
    \psi_{m}^{(0)*}
    +\psi_{ m}^{(2)*}
    \biggr)
    \Bigl(1-\frac{i}{2k\eta_i}\Bigr)
    e^{-ik\eta_i} \Biggr]\hat{a}_y^\dagger(\eta_i,-\bm{k})
    \nonumber\\
    &\hspace{1.5cm}
    +\psi_{p}^{(1)} e^{ik\eta_i}
    \hat{a}_A (\eta_i,\bm{k})
    +\psi_{m}^{(1)*} e^{-ik\eta_i}
    \hat{a}_A^\dagger(\eta_i,-\bm{k}),
    \label{y:bogoliubov1}
\end{align}
and the time evolution of annihilation operator of photon is expressed by the Bogoliubov transformation such as
\begin{align}
    &\hat{a}_A(\eta,\bm{k})=
    \Biggl( 
    \phi_{ p}^{(1)}
    \Bigl(1+\frac{i}{2k\eta_i}  \Bigr)
    e^{ik\eta_i} 
    + \phi_{ m}^{(1)*}
    \frac{i}{2k\eta_i}
    e^{-ik\eta_i} 
    \Biggr)\hat{a}_y(\eta_i,\bm{k})\nonumber\\
    &~~~~~~~~~~~
    +\Biggl(
    -\phi_{ p}^{(1)} \frac{i}{2k\eta_i}
    e^{ik\eta_i}
    +\phi_{ m}^{(1)*}
    \Bigl(1-\frac{i}{2k\eta_i}  \Bigr)
    e^{-ik\eta_i}
    \Biggr) \hat{a}_y^\dagger (\eta_i,-\bm{k})
    \nonumber\\
    &~~~~~~~~~~~
    +\Bigl(
    \phi_{p}^{(0)}+\phi_{p}^{(2)}
    \Bigr)e^{ik\eta_i} \hat{a}_{A}(\eta_i,\bm{k})
    +\Bigl(
    \phi_{m}^{(0)*}+\phi_{m}^{(2)*}
    \Bigr)e^{-ik\eta_i} \hat{a}_A^\dagger(\eta_i,-\bm{k}).
    \label{A:bogoliubov1}
\end{align}
These Bogoliubov transformations show the particle production during inflation and the mixing between graviton and photon.

It is useful to use a matrix form for later calculations. 
In fact, the Bogoliubov transofomation (\ref{y:bogoliubov1}) and (\ref{A:bogoliubov1}) and their conjugate can be accommodated into the simple $4\times 4$ matrix form $M$
\begin{eqnarray}
\begin{pmatrix}
a_y(\eta)\\
a_y^{\dagger}(\eta)\\
a_A(\eta)\\
a_A^{\dagger}(\eta)\\
\end{pmatrix}
=M
\begin{pmatrix}
a_y(\eta_i)\\
a_y^{\dagger}(\eta_i)\\
a_A(\eta_i)\\
a_A^{\dagger}(\eta_i)\\
\end{pmatrix}
=
\left\{
\begin{pmatrix}
A_{0}& 0\\
0  &D_{0}\\
\end{pmatrix} 
+
\begin{pmatrix}
0  &B_{1}\\
C_{1}&  0\\
\end{pmatrix} 
+
\begin{pmatrix}
A_{2}& 0\\
0  &   D_{2}\\
\end{pmatrix} 
\right\}
\begin{pmatrix}
a_y(\eta_i)\\
a_y^{\dagger}(\eta_i)\\
a_A(\eta_i)\\
a_A^{\dagger}(\eta_i)\\
\end{pmatrix}\ .
\nonumber
\hspace{-6mm}\\
\label{bogoliubov}
\end{eqnarray}
Here, the zeroth order Bogoliubov transformation consists of $2\times 2$ matrices $A_0$ and $D_0$ given by 
\begin{eqnarray}
A_0=
\begin{pmatrix}
K^* & -L^* \\
-L & K \\
\end{pmatrix}
\ ,\qquad
D_0=
\begin{pmatrix}
e^{ik\,(\eta-\eta_i)} & 0\\
0 & e^{-ik\,(\eta-\eta_i)} \\
\end{pmatrix} \ ,
\end{eqnarray}
where we defined 
\begin{eqnarray}
K&=&\left( 1+\frac{i}{2k\eta}\right)\left( 1-\frac{i}{2k\eta_i}\right)e^{ik(\eta-\eta_i)} 
-\frac{1}{4k^2\eta\eta_i }e^{-ik(\eta-\eta_i)}\ ,\\
L&=&-\frac{i}{2k\eta_i }\left( 1+\frac{i}{2k\eta}\right) e^{ik(\eta-\eta_i)} 
+\frac{i}{2k\eta }\left( 1+\frac{i}{2k\eta_i}\right) e^{-ik(\eta-\eta_i)} \ .
\end{eqnarray}
The first order Bogoliubov transofomation is written by $2\times 2$ matrices $B_1$ and $C_1$ such as
\begin{eqnarray}
B_1=
\begin{pmatrix}
e^{ik\eta_i}\psi_{p}^{(1)}   & e^{-ik\eta_i}\psi_{m}^{(1)*}\\
 e^{ik\eta_i}\psi_{m}^{(1)}& e^{-ik\eta_i}\psi_{p}^{(1)*} \\
\end{pmatrix}
\end{eqnarray}
and 
\begin{eqnarray}
C_1=
\begin{pmatrix}
\left( 1+\frac{i}{2k\eta_i}\right)e^{ik\eta_i}\phi_{p}^{(1)}+\frac{i}{2k\eta_i} e^{-ik\eta_i}\phi_{m}^{(1)*}  & \left( 1-\frac{i}{2k\eta_i}\right)e^{-ik\eta_i}\phi_{m}^{(1)*}-\frac{i}{2k\eta_i} e^{ik\eta_i}\phi_{p}^{(1)} \\
 \left( 1+\frac{i}{2k\eta_i}\right)e^{ik\eta_i}\phi_{m}^{(1)}+\frac{i}{2k\eta_i} e^{-ik\eta_i}\phi_{p}^{(1)*}& \left( 1-\frac{i}{2k\eta_i}\right)e^{-ik\eta_i}\phi_{p}^{(1)*}-\frac{i}{2k\eta_i} e^{ik\eta_i}\phi_{m}^{(1)} \\
\end{pmatrix}\ .
\end{eqnarray}
Finally, the second order Bogoliubov transformation $A_2$ and $D_2$ are
\begin{eqnarray}
A_2=
\begin{pmatrix}
\left( 1+\frac{i}{2k\eta_i}\right)e^{ik\eta_i}\psi_{p}^{(2)}+\frac{i}{2k\eta_i} e^{-ik\eta_i}\psi_{m}^{(2)*}  & \left( 1-\frac{i}{2k\eta_i}\right)e^{-ik\eta_i}\psi_{m}^{(2)*}-\frac{i}{2k\eta_i} e^{ik\eta_i}\psi_{p}^{(2)} \\
 \left( 1+\frac{i}{2k\eta_i}\right)e^{ik\eta_i}\psi_{m}^{(2)}+\frac{i}{2k\eta_i} e^{-ik\eta_i}\psi_{p}^{(2)*} & \left( 1-\frac{i}{2k\eta_i}\right)e^{-ik\eta_i}\psi_{p}^{(2)*}-\frac{i}{2k\eta_i} e^{ik\eta_i}\psi_{m}^{(2)}\\
\end{pmatrix}\ 
\end{eqnarray}
and
\begin{eqnarray}
D_2=
\begin{pmatrix}
e^{ik\eta_i}\phi_{p}^{(2)}   & e^{-ik\eta_i}\phi_{m}^{(2)*}\\
e^{ik\eta_i}\phi_{m}^{(2)} & e^{-ik\eta_i}\phi_{p}^{(2)*} \\
\end{pmatrix} \ .
\end{eqnarray}

\section{Time evolution of squeezing parameters}
In the previous section, we obtained the Bogoliubov transformation that mix the operators $\hat{a}_y(\eta)$, $\hat{a}_A(\eta)$
and their Hermitian conjugates $\hat{a}^\dagger_y(\eta)$, $\hat{a}^\dagger_A(\eta)$.
Note that the initial Bunch-Davies state is defined by
\begin{eqnarray}
\hat{a}_y(\eta_i,\bm{k})  |{\rm BD}\rangle= \hat{a}_A(\eta_i,\bm{k}) |{\rm BD}\rangle =0\,.
\label{BD}
\end{eqnarray}
In order to impose these conditions,
we need to invert the Bogoliubov transformations (\ref{y:bogoliubov1}) and (\ref{A:bogoliubov1})
into the form
\begin{eqnarray}
\hat{a}_y(\eta_i,\bm{k})
&=&\alpha_y\,\hat{a}_y (\eta,\bm{k})+ \beta_y\,\hat{a}_y^\dagger(\eta,-\bm{k})
  + \gamma_A\,\hat{a}_A(\eta,\bm{k})+ \delta_A\,\hat{a}_A^\dagger(\eta,-\bm{k})\,,
  \label{y:invert}\\
\hat{a}_A(\eta_i,\bm{k})
&=&\gamma_y\,\hat{a}_y(\eta,\bm{k})+ \delta_y\,\hat{a}_y^\dagger(\eta,-\bm{k}) 
  +\alpha_A\,\hat{a}_A (\eta,\bm{k})+ \beta_A\,\hat{a}_A^\dagger(\eta,-\bm{k})\,,
  \label{A:invert}
\end{eqnarray}
where $\alpha_y$, $\beta_y$, $\gamma_A$, $\delta_A$, $\gamma_y$, $\delta_y$, $\alpha_A$ and $\beta_A$ are the Bogoliubov coefficients and we will find these coefficients in the next subsection. 

\subsection{Inversion of the Bogoliubov transformation}

The matrix $M$ in Eq.~(\ref{bogoliubov}) can be expanded perturbatively as
\begin{eqnarray}
  M = M^{(0)}+M^{(1)}+M^{(2)}
  =M^{(0)}\left[ 1+ M^{(0)-1}M^{(1)}
  +M^{(0)-1}M^{(2)}\right] \ ,
\end{eqnarray}
where
\begin{eqnarray}
M^{(0)}=
\begin{pmatrix}
A_{0}& 0\\
0  &D_{0}\\
\end{pmatrix} 
\,,\qquad
M^{(1)}=
\begin{pmatrix}
0  &B_{1}\\
C_{1}&  0\\
\end{pmatrix} 
\,,\qquad
M^{(2)}=
\begin{pmatrix}
A_{2}& 0\\
0  &   D_{2}\\
\end{pmatrix}
\,.
\end{eqnarray}
Then the inverse of the $M$ is given by
\begin{eqnarray}
  M^{-1} 
  =\left[ 1- M^{(0)-1}M^{(1)}
  -M^{(0)-1}M^{(2)}
  + M^{(0)-1}M^{(1)}M^{(0)-1}M^{(1)} \right]M^{(0)-1}\ .
\end{eqnarray}
Using the above general formula, the inverse of the $M$ is obtained in the form
\begin{eqnarray}
M^{-1}=
\begin{pmatrix}
A_0^{-1}-A_0^{-1}A_2 A_0^{-1}
+A_0^{-1}B_{1}D_0^{-1}C_1A_0^{-1}& -A_0^{-1}B_{1}D_0^{-1}\\
-D_0^{-1}C_1 A_0^{-1}& D_0^{-1}-D_0^{-1}D_{2}D_0^{-1}
+D_0^{-1}C_{1}A_0^{-1}B_1D_0^{-1}
\label{inverseM}
\end{pmatrix} \,.
\nonumber
\hspace{-5mm}\\
\end{eqnarray}
We see that $A_0^{-1}$ and $D_0^{-1}$ are necessary to calculate the elements of the $M^{-1}$. They are given by
\begin{eqnarray}
A_0^{-1}=
\begin{pmatrix}
K & L^* \\
L & K^* \\
\end{pmatrix} 
\ , \qquad
D_0^{-1}=
\begin{pmatrix}
e^{-ik\,(\eta-\eta_i)} & 0\\
0 & e^{ik\,(\eta-\eta_i)} \\
\end{pmatrix}  \  .
\end{eqnarray}
From Eqs.~(\ref{y:invert}) and (\ref{A:invert}), the $M^{-1}$ is also written as
\begin{eqnarray}
M^{-1}=
\begin{pmatrix}
\alpha_y & \beta_y & \gamma_A & \delta_A \\
\beta_y^* & \alpha_y^* & \delta_A^* & \gamma_A^* \\
\gamma_y & \delta_y & \alpha_A & \beta_A \\
\delta_y^* & \gamma_y^* & \beta_A^* & \alpha_A^* \\
\end{pmatrix} \,,
\end{eqnarray}
where
\begin{eqnarray}
 && \alpha_y =\alpha_y^{(0)} +\alpha_y^{(2)} \,,\qquad
  \beta_y =\beta_y^{(0)} +\beta_y^{(2)} \,,\qquad
  \gamma_A =\gamma_A^{(1)} \,,\qquad
  \delta_A =\delta_A^{(1)} \,,
  \label{expand1}\\
&&   \alpha_A =\alpha_A^{(0)} +\alpha_A^{(2)} \,,\qquad
  \beta_A = \beta_A^{(2)} \,,\qquad
  \gamma_y =\gamma_y^{(1)} \,,\qquad
  \delta_y =\delta_y^{(1)} \ .
  \label{expand2}
\end{eqnarray}
The zeroth order elements are given by
\begin{eqnarray}
&& \alpha^{(0)}_y= \left( 1+\frac{i}{2k\eta}\right)\left( 1-\frac{i}{2k\eta_i}\right)e^{ik(\eta-\eta_i)} 
-\frac{1}{4k^2\eta\eta_i }e^{-ik(\eta-\eta_i)} \ ,\\
&& \beta^{(0)}_y = \frac{i}{2k\eta_i }\left( 1-\frac{i}{2k\eta}\right) e^{-ik(\eta-\eta_i)} 
-\frac{i}{2k\eta }\left( 1-\frac{i}{2k\eta_i}\right) e^{ik(\eta-\eta_i)} \ ,   \\
&& \alpha^{(0)}_A =e^{ik(\eta-\eta_i)} \ ,\qquad
 \beta^{(0)}_A = 0\ .
\end{eqnarray}
The first order elements are written as
\begin{eqnarray}
 \gamma^{(1)}_A &=& -\left( K\psi_{p}^{(1)}
+L^*\psi_{m}^{(1)}\right)e^{ik\eta}\ ,\\
 \delta^{(1)}_A &=& -\left(K\psi_{m}^{(1)*}
+L^*\psi_{p}^{(1)*}\right)
e^{-ik\eta} \ , \\
 \gamma^{(1)}_y &=& -K \left[\left( 1+\frac{i}{2k\eta_i}\right)e^{ik\eta}\phi_{p}^{(1)}
+\frac{i}{2k\eta_i} e^{ik(\eta-2\eta_i)}\phi_{m}^{(1)*} \right]\nonumber\\
&&-L \left[\left( 1-\frac{i}{2k\eta_i}\right)e^{ik(\eta-2\eta_i)}\phi_{m}^{(1)*}
-\frac{i}{2k\eta_i} e^{ik\eta}\phi_{p}^{(1)} \right]\ ,\\
 \delta^{(1)}_y  &=& -L^* \left[\left( 1+\frac{i}{2k\eta_i}\right)e^{ik\eta}\phi_{p}^{(1)}
+\frac{i}{2k\eta_i} e^{ik(\eta-2\eta_1)}\phi_{m}^{(1)*} \right]\nonumber\\
&&-K^* \left[\left( 1-\frac{i}{2k\eta_i}\right)e^{ik(\eta-2\eta_i)}\phi_{m}^{(1)*}
-\frac{i}{2k\eta_i} e^{ik\eta}\phi_{p}^{(1)} \right] \ .
\end{eqnarray}
The second order are 
\begin{eqnarray}
 \alpha^{(2)}_y &=& -K\left(KA_{11}+L^* A_{21}\right)
-L\left(KA_{12}+L^* A_{22}\right) \nonumber\\
&&+(C_{11} K+C_{12} L)\left(K\psi_{p}^{(1)}
+L^*\psi_{m}^{(1)}\right)e^{ik\eta}\nonumber\\
&&+(C_{21} K+C_{22} L)\left(K\psi_{m}^{(1)*}
+L^*\psi_{p}^{(1)*}\right)e^{-ik\eta}\ , \\
 \beta^{(2)}_y &=& -L^*\left(KA_{11}+L^* A_{21}\right)
-K^* \left(KA_{12}+L^* A_{22}\right) \nonumber\\
&&+(C_{11} L^* +C_{12} K^*)\left(K \psi_{p}^{(1)}
+L^*\psi_{m}^{(1)}\right)e^{ik\eta}\nonumber\\
&&+(C_{21} L^*+C_{22} K^*)\left(K\psi_{m}^{(1)*}
+L^*\psi_{p}^{(1)*}\right)e^{-ik\eta}\ ,\\ 
 \alpha^{(2)}_A &=& - e^{ik(2\eta-\eta_i)}\phi_{p}^{(2)} \nonumber\\
&& +(C_{11} K+C_{12} L)e^{ik(2\eta - \eta_i)}\psi_{p}^{(1)}
+(C_{11} L^* +C_{12} K^*)e^{ik(2\eta - \eta_i)}\psi_{m}^{(1)}
 \ , \\
 \beta^{(2)}_A &=&-e^{-ik\eta_i}\phi_{m}^{(2)*}\nonumber\\
&& +(C_{11} K+C_{12} L)  e^{-ik\eta_i }\psi_{m}^{(1)*}
+(C_{11} L^* +C_{12} K^*)e^{-ik\eta_i}\psi_{p}^{(1)*}\ ,
\end{eqnarray}
where we have defined
\begin{eqnarray}
A_{11}&=&\psi_{p}^{(2)}\left(1+\frac{i}{2k\eta_i}\right)e^{ik\eta_i}+\psi_{m}^{(2)*}\frac{i}{2k\eta_i}e^{-ik\eta_i}\,,\\
A_{12}&=&-\psi_{p}^{(2)}\frac{i}{2k\eta_i}e^{ik\eta_i}+\psi_{m}^{(2)*}\left(1-\frac{i}{2k\eta_i}\right)e^{-ik\eta_i}\,,\\
A_{21}&=&\psi_{p}^{(2)*}\frac{i}{2k\eta_i}e^{-ik\eta_i}+\psi_{m}^{(2)}\left(1+\frac{i}{2k\eta_i}\right)e^{ik\eta_i}\,,\\
A_{22}&=&\psi_{p}^{(2)*}\left(1-\frac{i}{2k\eta_i}\right)e^{-ik\eta_i}-\psi_{m}^{(2)}\frac{i}{2k\eta_i}e^{ik\eta_i}\,,
\end{eqnarray}
and
\begin{eqnarray}
C_{11}&=&\phi_{p}^{(1)}\left(1+\frac{i}{2k\eta_i}\right)e^{ik\eta_i}+\phi_{m}^{(1)*}\frac{i}{2k\eta_i}e^{-ik\eta_i}\,,\\
C_{12}&=&\phi_{m}^{(1)*}\left(1-\frac{i}{2k\eta_i}\right)e^{-ik\eta_i}-\phi_{p}^{(1)}\frac{i}{2k\eta_i}e^{ik\eta_i}\,,\\
C_{21}&=&\phi_{p}^{(1)*}\frac{i}{2k\eta_i}e^{-ik\eta_i}+\phi_{m}^{(1)}\left(1+\frac{i}{2k\eta_i}\right)e^{ik\eta_i}\,,\\
C_{22}&=&\phi_{p}^{(1)*}\left(1-\frac{i}{2k\eta_i}\right)e^{-ik\eta_i}-\phi_{m}^{(1)}\frac{i}{2k\eta_i}e^{ik\eta_i}\,.
\end{eqnarray}

\subsection{Squeezing operator}
In the previous subsection, we obtained the Bogoliubov coefficients of Eqs.~(\ref{y:invert}) and (\ref{A:invert}) up to the second order. If we apply the Eqs.~(\ref{y:invert}) and (\ref{A:invert}) to the definition of the Bunch-Davies vacuum~(\ref{BD}) and use the relations $[\hat{a}_y(\eta,{\bm k}),\hat{a}^\dag_y(\eta,-{\bm k}^\prime)]=\delta({\bm k}+{\bm k}^\prime)$\,, $[\hat{a}_A(\eta,{\bm k}),\hat{a}^\dag_A(\eta,-{\bm k}^\prime)]=\delta({\bm k}+{\bm k}^\prime)$ and $[\hat{a}_y(\eta,{\bm k}),\hat{a}_A(\eta,-{\bm k}^\prime)]=0$, the Bunch-Davies vacuum can be written by using squeezing parameters $\Lambda,\Xi$ and $\Omega$ such as 
\begin{eqnarray}
|{\rm BD}\rangle  = \prod_{k=-\infty}^{\infty}\exp\left[\frac{\Lambda}{2}\,
\hat{a}_y^\dag (\eta,\bm{k}) \hat{a}_y^\dag (\eta,-\bm{k})+\Xi\,\hat{a}_y^\dag (\eta,\bm{k}) \hat{a}_A^\dag (\eta,-\bm{k})
+\frac{\Omega}{2}\,\hat{a}_A^\dag (\eta,\bm{k}) \hat{a}_A^\dag (\eta,-\bm{k})\right]|0\rangle,\nonumber
\hspace{-5mm}\\
\end{eqnarray}
where $|0\rangle$ is the instantaneous vacuum defined by
\begin{eqnarray}
\hat{a}_y(\eta,{\bm k})  |0\rangle=\hat{a}_A(\eta,{\bm k}) |0\rangle =0 \,.
\end{eqnarray}
This describes a four mode squeezed state of pairs of graviton $y$ and photon $A$. In a different context, a four-mode squeezed state of two free massive scalar fields is discussed in~\cite{Albrecht:2014aga,Kanno:2015ewa}.
If we expand the exponential function in Taylor series, we find
\begin{eqnarray}
|{\rm BD}\rangle  =
\prod_{\bm k} \sum_{p\,,q\,,r=0}^{\infty}
 \frac{\Lambda^p\,\Xi^q\,\Omega^r}{2^{p+r}p!\,q!\,r!}  
 |p+q \rangle_{y,{\bm k}} \otimes |p \rangle_{y,-{\bm k}} \otimes |r \rangle_{A,{\bm k}} \otimes |q+r \rangle_{A,-{\bm k}}\,.
\end{eqnarray}
This is a four-mode squeezed state which consists of an infinite number of entangled particles in the ${\cal H}_{y,{\bm k}}\otimes{\cal H}_{y,{-\bm k}}\otimes{\cal H}_{A,{\bm k}}\otimes{\cal H}_{A,-{\bm k}}$ space. 
In particular, in the highly squeezing limit $\Lambda\,,\Xi\,,\Omega\rightarrow 1$, the Bunch-Davies vacuum becomes the maximally entangled state from the point of view of the instantaneous vacuum. 

Now we find the squeezing parameters.
The condition $\hat{a}_y(\eta_i,{\bm k})|{\rm BD}\rangle=0$ of Eq.~(\ref{BD}) yields
\begin{eqnarray}
    \alpha_y \Lambda +\beta_y +\gamma_A \Xi =0 \ , \qquad
    \alpha_y \Xi +\gamma_A \Omega +\delta_A  =0\,,
\end{eqnarray}
and another condition $\hat{a}_A(\eta_i,{\bm k}) |{\rm BD}\rangle=0$ of Eq.~(\ref{BD}) gives 
\begin{eqnarray}
    \alpha_A \Xi +\gamma_y \Lambda +\delta_y  =0 \ , \qquad
    \alpha_A \Omega +\beta_A  +\gamma_y \Xi =0\,.
\end{eqnarray}
Then, we obtain the three squeezing parameters $\Lambda,\Xi$ and $\Omega$ of the form
\begin{eqnarray}
    \Lambda= \frac{\gamma_A\delta_y -\beta_y \alpha_A}{\alpha_y \alpha_A -\gamma_y\gamma_A} \ , \qquad
    \Xi= \frac{\beta_y \gamma_y - \alpha_y \delta_y}{\alpha_y \alpha_A -\gamma_y\gamma_A} \ , \qquad
    \Omega= \frac{\gamma_y\delta_A -\beta_A \alpha_y}{\alpha_y \alpha_A -\gamma_y\gamma_A}\,.
    \label{squeezingparameters}
\end{eqnarray}
We have four relations for three parameters $\Lambda,\Xi$ and $\Omega$. The remaining relation is turned out to be guaranteed by the commutation relation:
\begin{eqnarray}
   [\hat{a}_y(\eta,{\bm k}) \ , \hat{a}_A(\eta,{\bm k}) ]
   = -\alpha_A\delta_A +\beta_A \gamma_A
   -\gamma_y \beta_y +\alpha_y \delta_y =0\,.
\end{eqnarray}
Thus, we find that Eq.~(\ref{squeezingparameters}) is the unique solution.
Since the Bogoliubov coefficients are given up to the second order as in Eqs.~(\ref{expand1}) and (\ref{expand2}),
the squeezing parameters can be expanded up to the second order such as
\begin{eqnarray}
  &&  \Lambda= -\frac{\beta_y^{(0)}}{\alpha_y^{(0)}} 
    \left[1-\frac{\alpha_y^{(2)}}{\alpha_y^{(0)}}
    + \frac{\beta_y^{(2)}}{\beta_y^{(0)}}
    + \frac{\gamma_y^{(1)}\gamma_A^{(1)}}{\alpha_y^{(0)}\alpha_A^{(0)}}
    -\frac{\gamma_A^{(1)}\delta_y^{(1)}}{\beta_y^{(0)}\alpha_A^{(0)}}
    \right]\ , \\
 &&   \Xi= \frac{\beta_y^{(0)}\gamma_y^{(1)}}{\alpha_y^{(0)}\alpha_A^{(0)}} 
    - \frac{\delta_y^{(1)}}{\alpha_A^{(0)}}\ , \\
 &&    \Omega= \frac{\delta_A^{(1)}\gamma_y^{(1)}}{\alpha_y^{(0)}\alpha_A^{(0)}}  
      - \frac{\beta_A^{(2)}}{\alpha_A^{(0)}}\ .
\end{eqnarray}
In this way, we obtained the squeezing parameters perturbatively up to the second order. We will discuss the behaviour of the squeezing of graviton $\Lambda$, the squeezing of mixing between graviton and photon $\Xi$ and the squeezing of photon $\Omega$ in the next section.

\subsection{Numerical and analytical results}

\begin{figure}[H]
\centering
 \includegraphics[width=\textwidth]{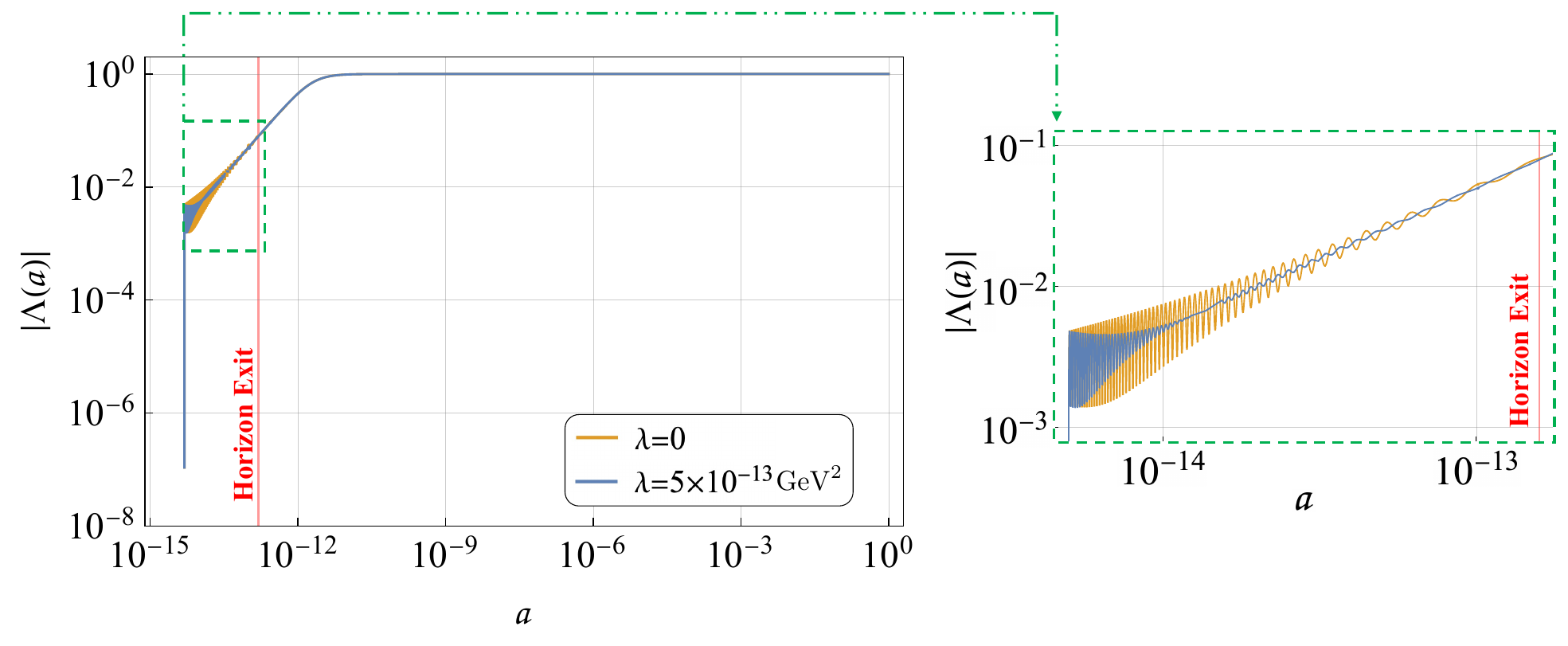}
 \renewcommand{\baselinestretch}{3}
 \caption{Squeezing parameter of graviton pair $\Lambda$ during inflation as a function of the scale factor $a(\eta)$. We set $\lambda=5\times10^{-13}{\rm GeV}^2$(blue line) and $\lambda=0~{\rm GeV}^2$(yellow line). 
 Other parameters are set as $k=10^2{\rm GeV}$, $H = 10^{14}{\rm GeV}$, $\eta_i=-2{\rm GeV}^{-1}$, $\eta_{f} =-10^{-14}{\rm GeV}^{-1}$, $a(\eta_i)=(2\times 10^{14})^{-1}$, and $a(\eta_{f})=1$. The red grid line shows the scale factor $a=1.59...\times10^{-13}$ at the time of horizon exit $\eta=-2\pi/k$.}
 \label{SqueezingA2}
 \end{figure}

\begin{figure}[H]
\centering
\includegraphics[width=0.55\textwidth]{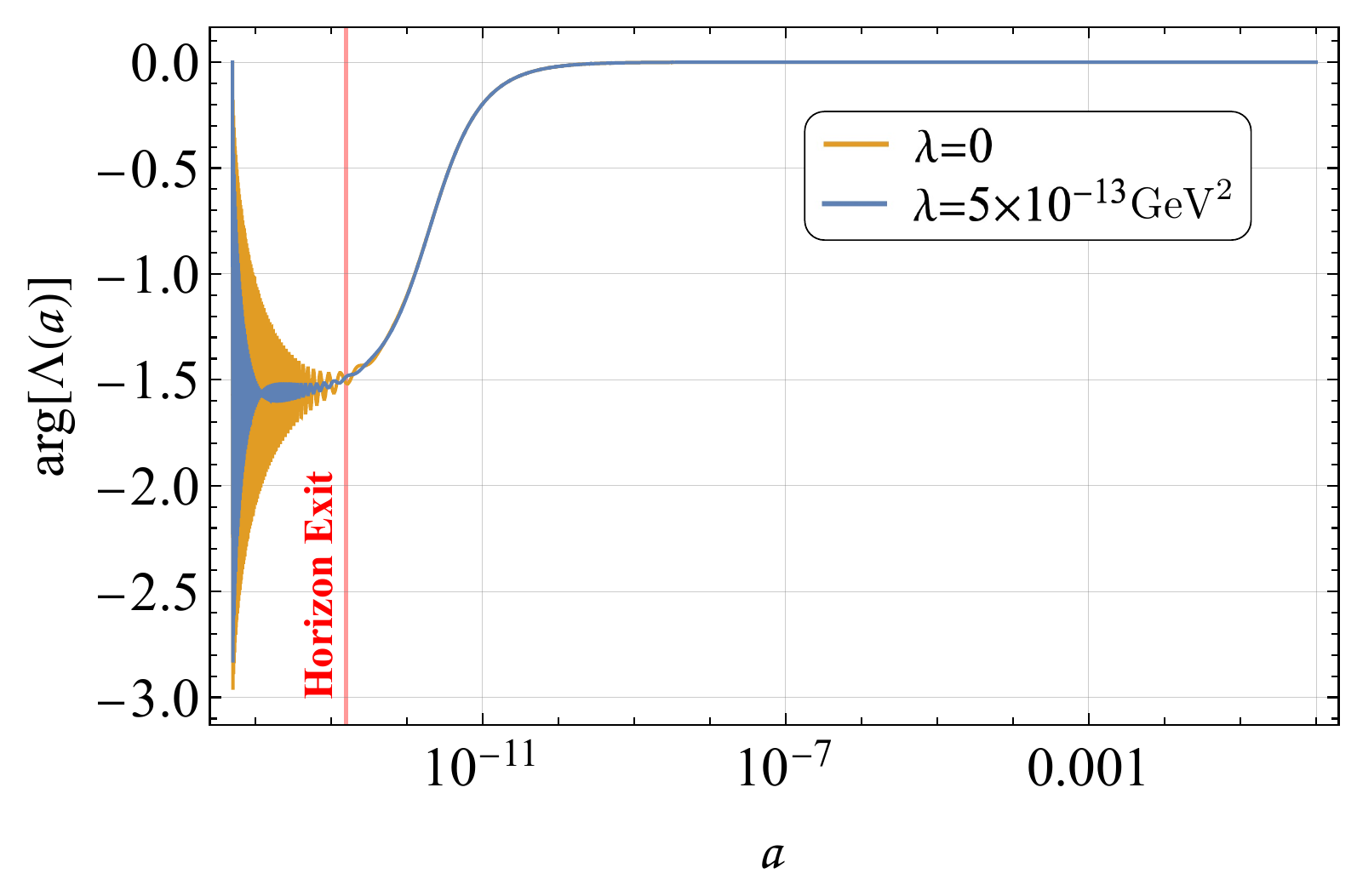}
\renewcommand{\baselinestretch}{1.2}
\caption{The phase of the squeezing  parameter of graviton pair $\Lambda(a)$ during inflation as function of the scale factor $a(\eta)$. We set $\lambda=5\times10^{-13}{\rm GeV}^2$(blue line) and $\lambda=0~{\rm GeV}^2$(yellow line). 
Other parameters are set as $k=10^2{\rm GeV}$, $H=10^{14}{\rm GeV}$, $\eta_i=-2{\rm GeV}^{-1}$, $\eta_f =-10^{-14}{\rm GeV}^{-1}$, $a(\eta_i)=(2\times 10^{14})^{-1}$, and $a(\eta_f)=1$. }
\label{PhaseA}
\end{figure}

\begin{figure}[H]
\centering
 \includegraphics[width=0.55\textwidth]{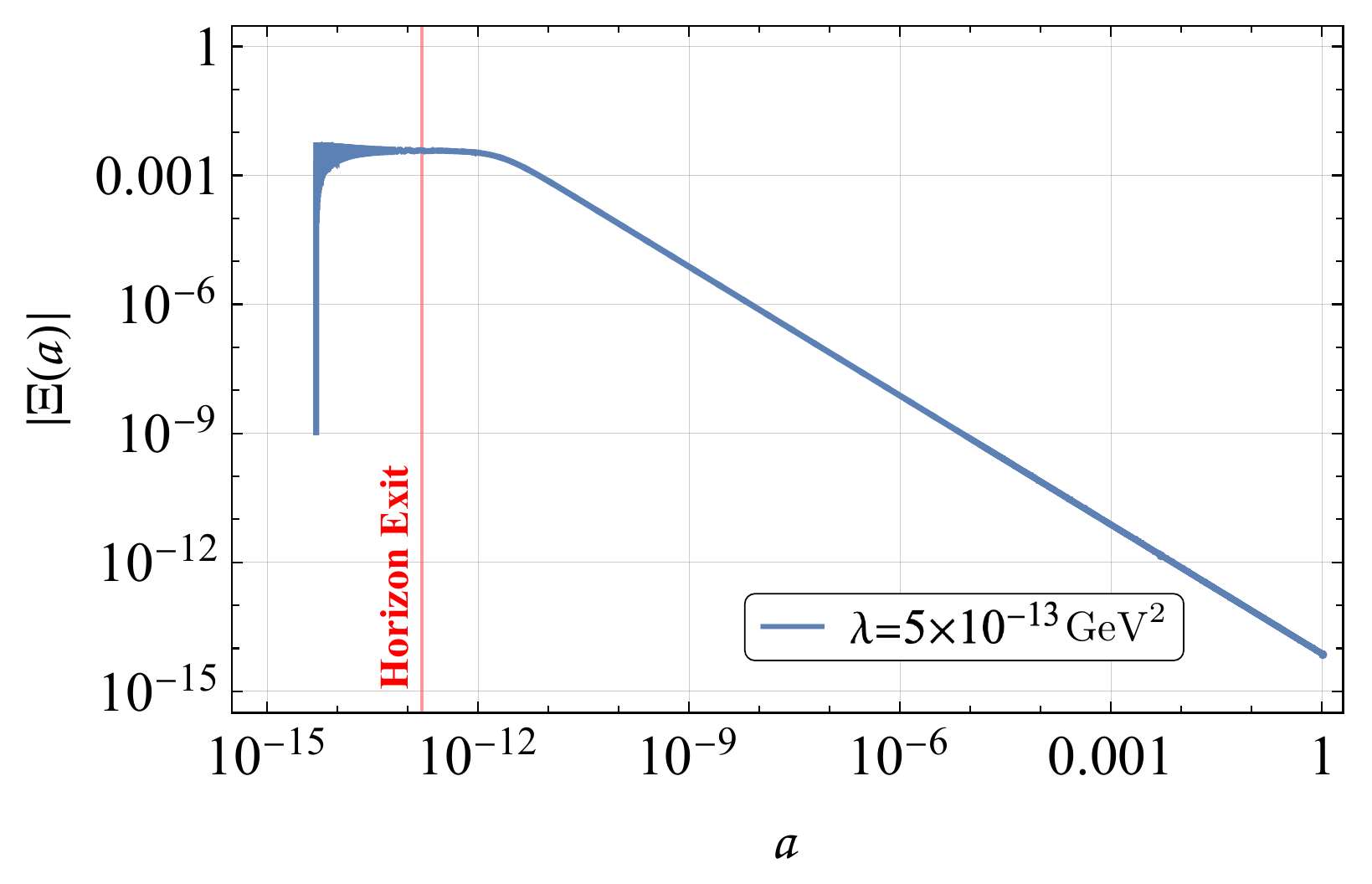}
  \renewcommand{\baselinestretch}{1.2}
 \caption{Squeezing parameter of graviton-photon pair $\Xi$ during inflation as a function of the scale factor $a$. We set  $\lambda=5\times10^{-13}{\rm GeV}^2$ (blue line). 
 Other parameters are set as $k=10^2{\rm GeV}$, $H=10^{14}{\rm GeV}$, $\eta_i=-2{\rm GeV}^{-1}$, $\eta_f =-10^{-14}{\rm GeV}^{-1}$, $a(\eta_i)=(2\times 10^{14})^{-1}$, and $a(\eta_f)=1$. }
 \label{SqueezingB2}
 \end{figure}

\begin{figure}[H]
\centering
\includegraphics[width=\textwidth]{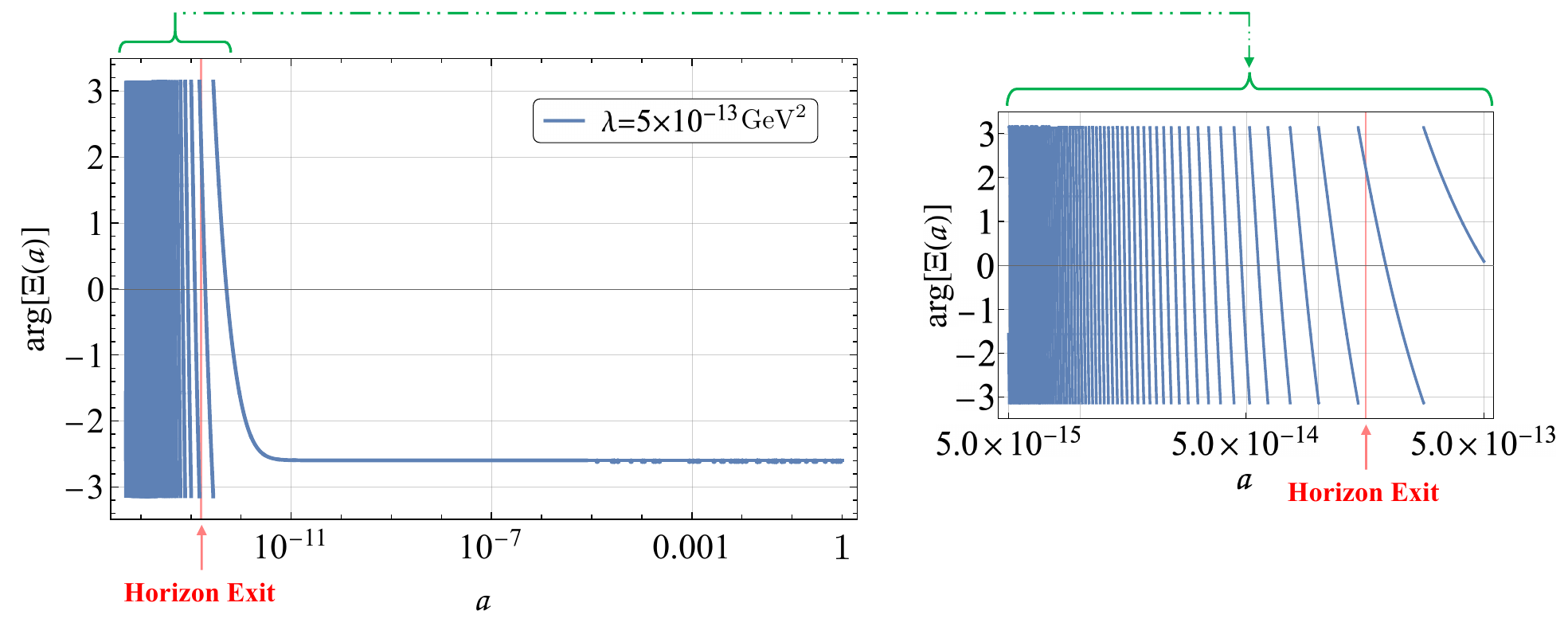}
\renewcommand{\baselinestretch}{1.2}
\caption{The phase of the squeezing parameter of photon pair $\Xi(a)$ during inflation as a function of the scale factor $a(\eta)$. We set $\lambda=5\times10^{-13}{\rm GeV}^2$, $k=10^2{\rm GeV}$, $H=10^{14}{\rm GeV}$, $\eta_i=-2{\rm GeV}^{-1}$, $\eta_f =-10^{-14}{\rm GeV}^{-1}$, $a(\eta_i)=(2\times 10^{14})^{-1}$, and $a(\eta_f)=1$. }
\label{PhaseB}
\end{figure}

\begin{figure}[H]
\centering
\includegraphics[width=0.55\textwidth]{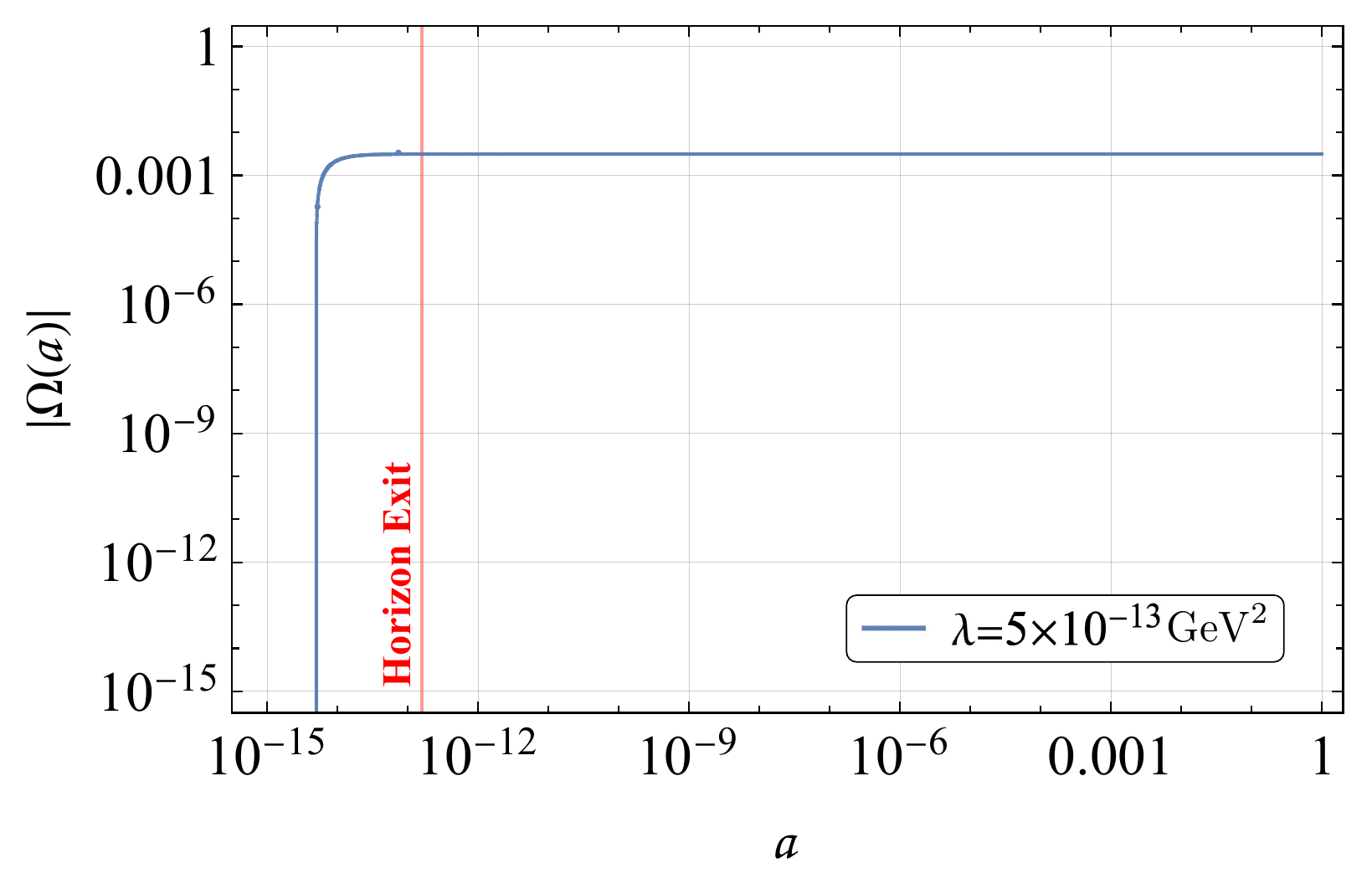}
\renewcommand{\baselinestretch}{1.2}
\caption{Squeezing  parameter of photon pair $\Omega$ during inflation as a function of the scale factor $a(\eta)$. We set  $\lambda=5\times10^{-13}{\rm GeV}^2$, $k=10^2{\rm GeV}$, $H=10^{14}{\rm GeV}$, $\eta_i=-2{\rm GeV}^{-1}$, $\eta_f =-10^{-14}{\rm GeV}^{-1}$, $a(\eta_i)=(2\times 10^{14})^{-1}$, and $a(\eta_f)=1$. }
\label{SqueezingC2}
\end{figure}

\begin{figure}[H]
\centering
\includegraphics[width=\textwidth]{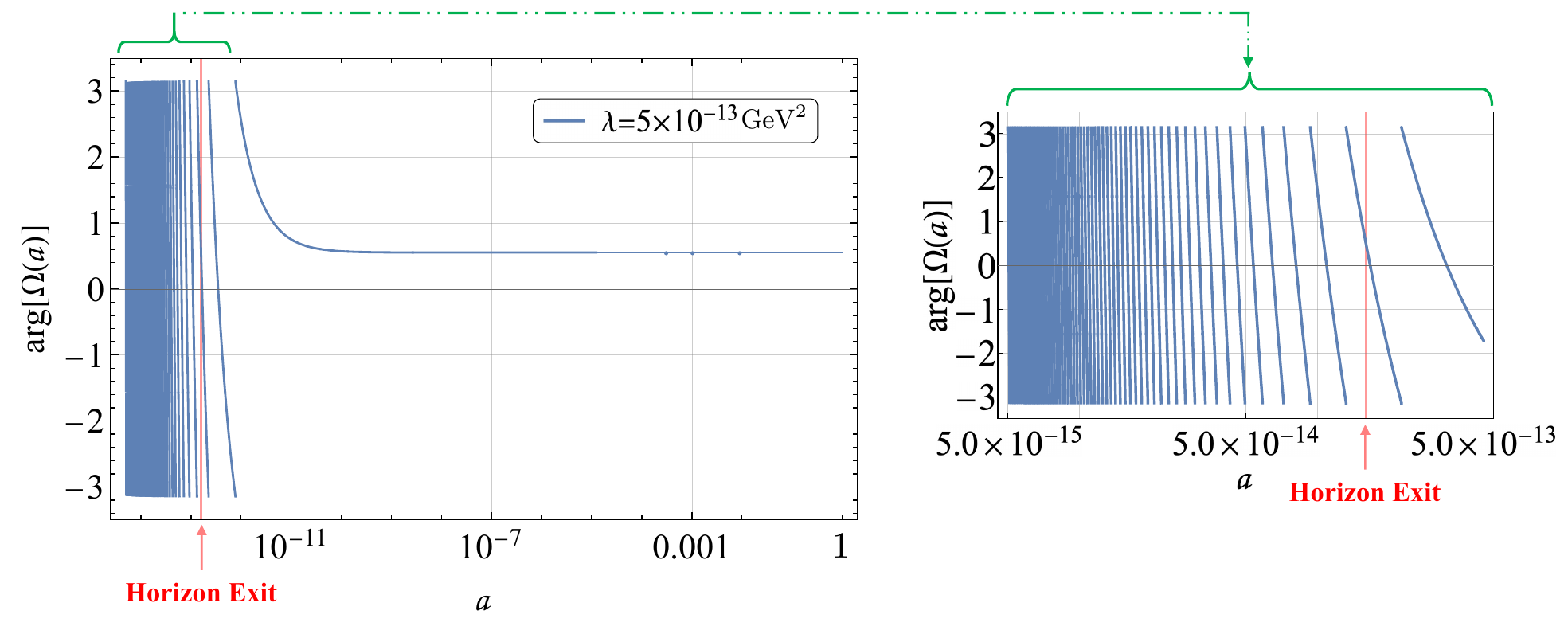}
\renewcommand{\baselinestretch}{1.2}
\caption{The phase of the squeezing  parameter of photon pair $\Omega$ during inflation as a function of the scale factor $a(\eta)$. We set $\lambda=5\times10^{-13}{\rm GeV}^2$, $k=10^2{\rm GeV}$, $H=10^{14}{\rm GeV}$, $\eta_i=-2{\rm GeV}^{-1}$, $\eta_f =-10^{-14}{\rm GeV}^{-1}$, $a(\eta_i)=(2\times 10^{14})^{-1}$, and $a(\eta_f)=1$. }
\label{PhaseC}
\end{figure}

The results of numerical calculations for the amplitude and the phase of the squeezing parameters $\Lambda$, $\Xi$, and $\Omega$ are plotted in FIGs.~\ref{SqueezingA2}, \ref{PhaseA}, \ref{SqueezingB2}, \ref{PhaseB}, \ref{SqueezingC2}, and \ref{PhaseC},  respectively, where we normalized the scale factor at the end of inflation as $a(\eta_f)=1$.
The evolution of the amplitude of $\Lambda$ in FIG. \ref{SqueezingA2} shows graviton is squeezed, that is, graviton pair production occurs  during inflation ($\eta<0$).
We see that sub-horizon modes oscillates rapidly and no graviton pair production seems to occur before horizon exit. In the presence of coupling with magnetic fields ($\lambda\neq 0$), the amplitude of oscillation is relatively small as represented by blue line. After horizon exit, the oscillation ceases and graviton pair production starts to occur and eventually $\Lambda$ becomes one. This means that almost maximum entangled pair of graviton are produced. This behavior does not change even for $\lambda\neq 0$.
The evolution of phase of $\Lambda$ is plotted in FIG. \ref{PhaseA}, in which we see the phase converges to zero. This is consistent with the result
in \cite{Polarski:1995jg}.
The time evolution of the amplitude of $\Xi$ in FIG. \ref{SqueezingB2} shows that one of pair of gravitons is converted to a photon and graviton-photon pair production occurs. 
We see that some amount of pair-production occurs when the mode leaves the horizon but the graviton-photon pair production decreases rapidly by the end of inflation.
The evolution of the phase of $\Xi$ plotted in FIG. \ref{PhaseB} is found to oscillate rapidly but eventually becomes constant after horizon exit. The similar behavior appears in the evolution of phase of $\Lambda$ in FIG.~{\ref{PhaseA}}.
However, the final
phase is found to depend on the initial condition in this case.
The time evolution of the amplitude of $\Omega$ in FIG. \ref{SqueezingC2} tells us that photon is squeezed, that is, graviton pair production is converged to photon pair production. 
Interestingly, photon pair production occurs rapidly only at the initial time and no more production occurs after that. 
The behavior of the phase evolution of $\Omega$ 
in FIG.\ref{PhaseC} is similar to that of $\Xi$.

Now, we investigate the behavior of those squeezing parameters for $k\eta\ll 1$ and $k\eta_i \gg 1$ analytically.  The leading and sub-leading terms of $\Lambda$ and $\Xi$ can be calculated as
\begin{align}
\Lambda=1+   {\cal O} \left(\frac{\lambda^2 H^2\eta_i^2 }{k^4  }\right)\,,\qquad
\Xi=0+ {\cal O} \left(\frac{\lambda H \eta}{k^2   }\right)\,.
\label{lambdaxi}
\end{align}
We find that sub-leading terms of $\Lambda$ and $\Xi$ are negligibly small near the end of inflation and which is consistent with the numerical results in  FIGs.~\ref{SqueezingA2} and \ref{SqueezingB2}. This result tells us that the conversion from graviton pair production to graviton-photon pair production is hard to occur.
For the squeezing parameter $\Omega$, we find
\begin{align}
\Omega= i e^{2ik \eta_i} \frac{5\lambda^2H^2 \eta_i^3 }{32k^3}\,.
\label{main}
\end{align}
If we use the numerical values $\lambda=5\times10^{-13}\,{\rm GeV}^2$,  
 $k=10^2\,{\rm GeV}$, $H=10^{14}\,{\rm GeV}$, and $\eta_i=-2\,{\rm GeV}^{-1}$
, we find $|\Omega| \sim 0.003$ and which agrees with the numerical result in FIG. \ref{SqueezingC2}. Thus only small amount of conversion from graviton pair production to photon pair production occurs at the end of inflation.
These results support the validity of our iterative method to derive squeezing parameters.

Let us discuss implications of our numerical and analytical results. If the squeezing of graviton decreases as time evolves, it implies that the decoherence of graviton occurs. 
However, we found that 
the squeezing parameter of graviton pair increases and becomes $\Lambda\rightarrow 1$, so it seems that the decoherence is hard to occur.
This behavior can be understood as follows.
Since the effective coupling $\lambda H\eta$ between graviton and photon in Eqs.~(\ref{eom:graviton}) and (\ref{eom:photon}) decreases and eventually becomes negligible as $\eta\rightarrow 0$ during inflation, practically graviton-photon conversion stops.
Even after the graviton-photon conversion stops, the squeezing process of graviton pair continues as time evolves during inflation, so the squeezing of graviton pair $\Lambda$ continues to grow irrespective of the presence of the magnetic field as shown in FIG.~\ref{SqueezingA2}. 
Next, from FIG.~\ref{SqueezingB2}, we see the squeezing parameter of graviton-photon pair vanishes $\Xi\rightarrow 0$ as time evolves. This is consistent with Eq.~(\ref{lambdaxi}).
This is because the graviton-photon pair production is possible only in the presence of magnetic fields
due to spin conservation. In our setup, however, the energy density of the background magnetic field decreases proportional to $a(\eta)^{-4}$ as the universe expands. 
Hence, the rapid decay of magnetic fields lead to the rapid decay of $\Xi$.
Finally, we consider the evolution of the squeezing parameter of photon pair $\Omega$. By using the coupling constant $\lambda \simeq Bk /M_{\rm pl}$ in Eq.~(\ref{coupling}) and the scale factor at the initial time $a_i \equiv -1/(H\eta_i)$, the $\Omega$ reads
\begin{eqnarray}
\Omega\simeq  \frac{B^2}{a_i^4 M_{\rm pl}^2 H^2}
\frac{1}{k\eta_i} \ .
\end{eqnarray}
The first factor is the ratio of the energy density of the background magnetic field at the time $\eta_i$
to that of the inflaton field. The second factor is the ratio of the mode of graviton to the Hubble radius. In order to have inflation, the energy density of the magnetic field has to be smaller than  that of inflaton fields, that is, $B^2/a_i^4\ll M_{\rm pl}^2H^2$. And all modes of graviton is inside horizon initially, that is,  $1/k\ll \eta_i$. Hence, the $\Omega$
never exceeds unity after time evolution, which is consistent with FIG.~\ref{SqueezingC2}. 
Moreover, since graviton-photon conversion stops, 
the squeezing of photon pair $\Omega$ converges to a constant value as shown in FIG.~\ref{SqueezingC2}.

\section{Conclusion}

The relic gravitons are expected to be squeezed during inflation. In that case, quantum noise induced by them can be significantly enhanced in current interferometers. 
However, we need to properly take into account the decoherence of the relic gravitons during cosmic history.
As a first modest step in this direction, we assumed the presence of a sizable magnetic field at the beginning of inflation. If the squeezing of graviton decreases as time evolves, it implies that the decoherence of graviton occurs.
So, we studied the conversion processes of the squeezed gravitons into photons during inflation in the case of minimal coupling between gravitons and photons.
We solved the dynamical evolution of a coupled system of graviton and photon 
perturbatively. We numerically plotted
the squeezing parameters for the system of graviton and photon. 
FIG.~\ref{SqueezingA2} showed that magnetic fields do not affect the graviton squeezing parameter. 
In FIG.~\ref{SqueezingB2},
we numerically checked the parameter of squeezed graviton-photon pair $\Xi$ and found that the $\Xi$ rapidly decays at the end of inflation.
This fact was confirmed analytically in Eq.~(\ref{lambdaxi}). 
We found that the rapid decay of the initial presence of magnetic fields leads to the rapid decay of the $\Xi$.
In FIG.~\ref{SqueezingC2}, 
we depicted the squeezing parameter of the photon. It turned out that the amount of squeezed photon produced by the conversion was tiny. 
We derived an analytic formula for the
squeezing parameter of photons $\Omega$
and found that the degree of squeezing is at a few percent at most.

Since we found that gravitons are robust against the decoherence caused by the cosmological magnetic field, we could expect to find squeezed relic gravitons through  quantum noise induced by them in interferometers~\cite{Parikh:2020nrd,Kanno:2020usf,Parikh:2020kfh,Parikh:2020fhy,Kanno:2021gpt}.
We should note that the analysis in our paper can also be applicable to the dark magnetic field models~\cite{Masaki:2018eut} based on the dark photon scenario~\cite{Caputo:2021eaa}.

There are several directions to be pursued. 
It would be intriguing to follow up the evolution of the squeezed relic gravitons up to the radiation-dominated 
and matter-dominated eras. If we could show the absence of decoherence of the squeezed relic gravitons, the robustness of them would be proved.
It would also be interesting to study the case that the primordial magnetic fields persist against the cosmic no-hair theorem during inflation~\cite{Kanno:2009ei}.
On top of gravitons, the squeezing occurs for the light axion dark matter fields~\cite{Kuss:2021gig,Kanno:2021vwu}.
The decoherence of axion fields  due to magnetic fields can be discussed in a similar way.
We leave these issues for future work.

\section*{Acknowledgments}
S.\ K. was supported by the Japan Society for the Promotion of Science (JSPS) KAKENHI Grant Number JP22K03621.
J.\ S. was in part supported by JSPS KAKENHI Grant Numbers JP17H02894, JP17K18778, JP20H01902, JP22H01220.
K.\ U. was supported by the Japan Society for the Promotion of Science (JSPS) KAKENHI Grant Number 20J22946.

\printbibliography
\end{document}